\documentclass[traditabstract]{aa}
\usepackage{graphicx}
\usepackage{txfonts}
\usepackage{longtable,lscape}
\usepackage{natbib}
\usepackage{hyperref}

\bibpunct{(}{)}{;}{a}{}{,}

\begin{document}
\title{Kinematic properties of the field elliptical NGC
  7507\thanks{Based on observations taken at the Gemini Observatory,
    which is operated by the Association of Universities for Research
    in Astronomy, Inc., under a cooperative agreement with the NSF on
    behalf of the Gemini partnership: the National Science Foundation
    (United States), the Science and Technology Facilities Council
    (United Kingdom), the National Research Council (Canada), CONICYT
    (Chile), the Australian Research Council (Australia), Minist\'erio
    da Ci\^encia e Tecnologia (Brazil) and SECYT (Argentina).}}

\author{ 
R. Salinas\thanks{Current address: Finnish Centre for
    Astronomy with ESO (FINCA), University of Turku, V\"ais\"al\"antie
    20, FI-21500 Piikki\"o, Finland; rcsave@utu.fi} \inst{1,2}
\and 
T. Richtler  \inst{1}
\and 
L. P. Bassino \inst{3}
\and 
A. J. Romanowsky  \inst{4}
\and 
Y. Schuberth \inst{5}
}

\institute{
Departamento de Astronom\'{\i}a, 
Universidad de Concepci\'on, 
Concepci\'on, Chile 
\and
European Southern Observatory, 
Alonso de C\'ordova 3107, Santiago, Chile
\and
Facultad de Ciencias Astron\'omicas y Geof\'{\i}sicas de la Universidad Nacional de La Plata and 
Instituto de Astrof\'isica de la Plata
(CCT La Plata-CONICET-UNLP),
Paseo del Bosque S/N, 1900 La Plata, Argentina
\and
UCO/Lick Observatory, University of California, 
Santa Cruz, CA 95064, USA
\and             
Argelander Institut f\"ur Astronomie, 
Auf dem H\"ugel 71, 53121 Bonn, Germany 
}

\date{Received  / Accepted }

%%%%%%%%%%%%%%%%%%%%%%%%%%%%%%%%%%%%%%%%%%%%%%%%%%%%%%%%%%%%%%%%%%%%%%
\abstract{The dark matter (DM) halos of field elliptical galaxies have
  not been well studied and their properties appear controversial in
  the literature. While some galaxies appear to be nearly devoid of
  DM, others show clear evidence of its presence. Furthermore,
  modified Netonian dynamics (MOND), which has been found to have
  predictive power in the domain of disk galaxies, has not yet been
  investigated for isolated elliptical galaxies. We study the
  kinematics of the isolated elliptical NGC 7507, which has been
  claimed as a clear example of DM presence in early-type galaxies. We
  obtained major and minor axis long-slit spectroscopy of NGC 7507
  using the Gemini South telescope and deep imaging in Kron-Cousins
  $R$ and Washington $C$ using the CTIO/MOSAIC camera. Mean
  velocities, velocity dispersion and higher order moments of the
  velocity distribution are measured out to $\sim$90\arcsec. The
  galaxy, although almost circular, has significant rotation along the
  minor axis and a rapidly declining velocity dispersion along both
  axes. The velocity dispersion profile is modeled in the context of a
  spherical Jeans analysis. Models without DM provide an excellent
  representation of the data with a mass-to-light ratio ($M/L$) of 3.1
  ($R$-band). The most massive Navarro-Frenk-White (NFW) halo the data
  allow has a virial mass of only $3.9^{+3.1}_{-2.1}\times10^{11}$
  M$_{\sun}$, although the data are more consistent with models that
  have a slight radial anisotropy, which implies the galaxy has an
  even lower DM halo mass of $2.2^{+2.0}_{-1.2}\times10^{11}$
  M$_{\sun}$. Modeling of the $h_4$ Gauss-Hermite coefficient is
  inconclusive but seems to be consistent with mild radial
  anisotropy. A cored logarithmic DM halo with parameters $r_0$\,=\,7
  kpc and $v_0$\,=\,100 km s$^{-1}$ can also reproduce the observed
  velocity dispersion profile. The MOND predictions overestimate the
  velocity dispersion. In conclusion, we cannot easily reproduce the
  previous findings of a predominance of DM in NGC 7507 within a
  simple spherical model. DM may be present, but only in conjunction
  with a strong radial anisotropy, for which there are some
  indications.}

\keywords{Galaxies: individual: NGC 7507 -- Galaxies: kinematics and
  dynamics}

\maketitle

%%%%%%%%%%%%%%%%%%%%%%%%%%%%%%%%%%%%%%%%%%%%%%%%%%%%%%%%%%%%%%%%%%%%%
\section{Introduction}
\label{sec:intro}
The existence of DM in spiral galaxies (or alternatively a law of
gravitation that differs from the Newtonian law) remains
undisputed. However, the same cannot be said conclusively for
\emph{all} elliptical galaxies. Central galaxies in galaxy clusters
such as NGC 6166 \citep{kelson02}, NGC 1399
\citep{richtler08,schuberth10}, and NGC 3311 \citep{richtler11} are
embedded in massive dark halos as is the case of bright non-central
cluster ellipticals such as NGC 4636 and NGC 4374 in Fornax
\citep{schuberth06,napolitano10}. In constrast elliptical galaxies
with lower luminosities in sparser environments (NGC 821, 3379, 4494,
4697), whose mass profiles have been studied using planetary nebulae
(PNe) kinematics, appear to show only weak evidence of DM \citep[][but
see
\citealt{weijmans09}]{romanowsky03,douglas07,napolitano09,mendez09,teodorescu10},
although the case of NGC 3379 may be consistent with a normal cold
DM halo if a very strong radial anisotropy is present
\citep{delorenzi09}.

The comparison of field, and possibly isolated, elliptical galaxies to
those in clusters is of great interest. Different formation mechanisms
may have led to differences in their dark halo structure, but
observationally few properties have been reliably determined. Isolated
galaxies may also be good test objects of the Modified Newtonian
Dynamics \citep[MOND,][]{milgrom83,sanders02}, since complications
arising from the ``external field effect''\citep{milgrom83,wu07}
should be minimal.

The galaxy NGC 7507 is an elliptical \citep[E0,][]{desouza04} of
$M_V$=$-21.6$, with a distance modulus of 31.83$\pm$0.17 based on
surface brightness fluctuations \citep[23.22$\pm$1.8 Mpc;][ corrected
by $-0.16$ mag following \citealt{jensen03}]{tonry01}; we assume this
distance throughout the paper, which implies a scale of 112.5 pc
arcsec$^{-1}$. The galaxy is fairly isolated, forming a probable pair
with the SBb galaxy NGC 7513, which lies at a projected distance of 18
arcminutes ($\sim$ 130 kpc), but displaying no signs of recent
interactions \citep{tal09}. Due to its proximity and regular shape,
its dynamics have been well studied leading to strikingly different
conclusions. \citet{bertin94} found no evidence of a DM halo,
using major-axis kinematical data out to 66\arcsec. \citet[][hereafter
K+00]{kronawitter00} performed dynamical analyses of 21 round
elliptical galaxies, including NGC 7507. They concluded that models
based on luminous matter could only be ruled out with 95\% confidence
for only three galaxies, including NGC 7507. Contradicting this
finding, \citet{magorrian01} adopted the kinematics of
\citet{bertin94} and photometry of \citet{franx89a} to establish that
NGC 7507 has a constant mass-to-light ratio ($M/L$) profile, i.e. that
it does not contain a dark halo.  \defcitealias{kronawitter00}{K+00}

The most significant difference between these studies appears to be
the observational data used and not the modeling approach, which,
although different, should not give strongly discrepant results. We
decided to obtain new, deeper long-slit spectroscopy using Gemini
South. Eight-meter class telescopes have been scarcely used to measure
the long-slit kinematics of the unresolved stellar components of
elliptical galaxies \citep[e.g.][and see \citealt{samurovic05} for a
review of the work made with 4m class
telescopes]{thomas07,forestell10,pu10}. Much effort has been put into
kinematic studies of PNe \citep[e.g.][]{mendez09} and globular
clusters \citep[e.g.][]{romanowsky09b,schuberth10}, which can be
observed out to much larger galactocentric radii, but provide weaker
constraints of their orbital anisotropies.

The paper is organized as follows: we present our observations and
reduction procedures in Sect. \ref{sec:obs}. The photometric
properties of the galaxy together with a description of its kinematics
are given in Sect. \ref{results}. The dynamical modeling is presented
in Sect. \ref{sec:models}, while the discussion and conclusions can be
seen in Sects. \ref{sec:disc} and \ref{sec:conclusions}, respectively.

%%%%%%%%%%%%%%%%%%%%%%%%%%%%%%%%%%%%%%%%%%%%%%%%%%%%%%%%%%%%%%%%%%%%
\section{Observations and data reduction}
\label{sec:obs}
\subsection{CTIO/MOSAIC photometry and photometric calibration}

The images used in this study were obtained with the MOSAIC II camera
(an 8 CCDs mosaic imager, 16 amplifiers) mounted at the prime focus of
the 4-m Blanco telescope at the Cerro Tololo Inter-American
Observatory (CTIO, Chile).  Observations were carried out during 2005
August 5 -- 6, where the first night was ``useful'' and the second one
``photometric'', according to the CTIO reports. One pixel of the
MOSAIC wide-field camera subtends 0.27 arcsec on the sky, which
corresponds to a field of 36\arcmin $\times$ 36\arcmin,
i.e. approximately 230 $\times$ 230 kpc$^2$ at the distance of
NGC\,7507.

All fields were imaged in Washington $C$
($\lambda_{\mathrm{central}}\!=\!3850\,\AA$) and Kron-Cousins $R$
($\lambda_{\mathrm{central}}\!=\!6440\,\AA$) filters. The $R$ filter
was used instead of the original Washington $T_1$, as
\citet{geisler96} has shown that the Kron-Cousins $R$ filter is more
efficient than $T_1$, and that $R$ and $T1$ magnitudes are closely
related, with just a very small colour term and zero-point difference
($R\!-\!T1\!\sim\!-0.02$).

To fill in the gaps between the eight individual MOSAIC
chips, the data were dithered taking four images in $R$ with exposure
times of 720 s each, and seven images in $C$ with exposures of 1800 s
each. To avoid saturation at the galaxy core, additional shorter
exposures of 60 s in $R$ and of 300 s in $C$ were also obtained. Sky
flats were obtained at the beginning and end of each night.

The MOSAIC data reduction was performed using tasks from the
\verb+MSCRED+ package within IRAF. Several tasks were employed to
correct for the variable pixel scale across the CCD, which might cause
a 4 per cent variability in the point-sources brightness (from the
center to the corners).  The final combined $R$ image still shows
sensitivity variations up to 1.4 per cent, and the final $C$ image up
to 3.0 per cent (peak-to-peak).  The seeing in these final images is
1.1$\arcsec$ on the $R$ frame, and 1.3 $\arcsec$ on the $C$ image.

The photometric calibration was performed with three fields of
standard stars selected from the list of \citet{geisler96}, each
containing 7 to 10 standard stars. For these fields, $C$ and $R$
images were obtained in each night, covering an airmass range 1.1 --
2.3. Finally, we used a single set of calibration equations for both
nights.

The equations read
\begin{eqnarray*}
T1 = R_{\mathrm{inst}} + (0.89 \pm 0.01) - (0.23 \pm 0.01)X_R \\+ (0.016 \pm
0.003)(C-T1),\\
C = C_{\mathrm{inst}} + (1.55 \pm 0.01) - (0.30 \pm 0.01)X_C \\+ (0.069 \pm 0.003)(C-T1),
\end{eqnarray*}
where $C_{\mathrm{inst}}$ and $R_{\mathrm{inst}}$ are instrumental
magnitudes and $X_C$ and $X_R$ are the airmasses.

The absolute photometric quality of the $C$-photometry is dubious,
hence it is only used for differential analyses.  The $R$-photometry
is of high quality, as the comparison of our results with previous
photometric data sets of the galaxy reveals. The mean difference from
\citet{franx89a} is 0.08 mag (Franx et al. is brighter). We also
compare with photoelectric aperture measurements given by
\citet{poulain94}. In this case, we find that our profile is brighter
by 0.15 mag.

\subsection{Surface photometry}

The surface photometry of the galaxy was perfomed using the
\verb+IRAF/ELLIPSE+ task. The shallower images were used to determine
the surface brightness profile for the inner $\sim$10$\arcsec$ in the
$C$ and $R$ profiles, which were saturated in the longer
exposures. The sky brightness in the shallow images was determined by
taking the median value of the sky in several $15\times15$ pixel boxes
of empty regions further than 5$\arcmin$ from the galaxy center. Since
an accurate sky determination is critical in shaping the light profile
in the outer parts of a galaxy, a different procedure, similar to the
one devised by \citet{dirsch03}, was adopted for the deep
images. First, \verb+DAOPHOT+ \citep{stetson87} was applied to all
point sources present in the field.  The local sky brightness of each
source was determined using the median of the sky value of the pixels
between 15 and 25 pixels from each source center. Deviant pixels were
rejected using a three-sigma clipping procedure. All the sky values
were then ordered according to their distance from the center of NGC
7507. To exclude overestimated sky values (of point sources close to
nearby galaxies, bright stars or chip defects), a new iterative
3-sigma clipping was applied to the sky values within radial bins of
100 objects, this time using robust location and scale bi-weight
estimators \citep{beers90}. At distances larger than 5$\arcmin$ from
the center of the galaxy, the value of the sky brightness fluctuates
by less than 0.5\%. The final value for the sky was then taken as the
mean of the sky values between 10$\arcmin$ and 12$\arcmin$
($\sim$78--98 kpc) from the galaxy center and the $R$ sky brightness
in mag/arcsec$^2$ was found to be $20.90$. The photometry of the
galaxy can be seen in Table \ref{table:photometry}.

\subsection{Gemini/GMOS spectroscopy}

Long-slit spectra of NGC 7507 were obtained using the Gemini
Multi-Object Spectrograph (GMOS) on Gemini South, Cerro Pach\'on,
Chile, in queue mode during the nights of September 19 and 21, 2004
(program GS-2004B-Q-75). The grism was the B600+\_G5323 which, in
conjunction with a 1'' slit width, gives a resolution of
$\sim$4.7$\AA$ FWHM. Three consecutive exposures of 1800s each were
taken for a position angle of 90 $\degr$, which is close to the major
axis ($PA\!=\!105\degr$), and for $PA\!=\!0\degr$, which is close to
the minor axis ($PA\!=\!15\degr$). For convenience, these two
positions are referred to hereafter as the major and minor axes,
respectively.  Standard reduction steps (bias subtraction, flat
fielding, detectors mosaicing, wavelength calibration) were performed
with the Gemini IRAF package (v 1.9.1). Typical wavelength residuals
were $\sim$0.05\,\AA. No cosmic ray cleaning was attempted at this
stage because it introduced undesired noise, especially around bright
sky lines. Since copper-argon arc spectra were taken during daytime
and not immediately either after or before the science exposures, a
correction of the zero point of the wavelength solution was necessary
and was performed by measuring the position of the bright sky line at
5577.34\,\AA. This correction typically amounted to 0.1\,\AA. After
that, the three spectra for each position angle were averaged. Sky
subtraction was done using the \verb+IRAF+ task \verb+background+, by
averaging 50 rows on opposite sides of each spectra, at around
4.5$\arcmin$ from the galaxy center for the major axis and
$\sim4\arcmin$ for the minor axis. The two-dimensional (2D) spectra
were spatially rebinned to achieve a constant $S/N \sim 50/$pixel,
which is highly desirable when measuring the higher order moments of
the line-of-sight velocity distribution (LOSVD). The 4800--5500\,\AA\,
wavelength range, which contains strong absorption features as
H$\beta$, Mg $b$, and Fe lines around 5325\,\AA, was selected for the
kinematical analysis.

Mean velocity, velocity dispersion and higher order moment profiles
were measured for both long slits by using the Gauss-Hermite pixel
fitting code \citep[][hereafter vdM code]{vdm94} and the penalized
pixel fitting (pPXF) method developed by \citet{cappellari04}. Both
approaches have the advantage of performing their calculations in
pixel space instead of Fourier space, thus allowing a more precise
masking of undesired zones in the spectra (emission lines, bad
columns, etc). Both methods also parametrize the LOSVD in terms of a
Gauss-Hermite series \citep{gerhard93,vdm93}. The former algorithm
measures the mean velocity and the velocity dispersion, and, with
these values fixed, fits the higher order terms of the velocity
distribution, parametrized by the Gauss-Hermite coefficients $h_3$ and
$h_4$. The latter algorithm fits all parameters simultaneously, but
can bias the solution to a Gaussian shape when the higher order terms
are not sufficiently constrained by the data \citep{cappellari04}.  An
advantage of the pPXF method over the vdM code is that it can perform
the fitting process with a linear combination of template spectra,
minimizing the effect of template mismatching.
 
A subset of 22 old and metal-rich single stellar population models
from \citet{vazdekis10} based on the MILES stellar library
\citep{sanchez06} were used as velocity templates for our analysis
using the vdM code.  These templates were logarithmically rebinned to
the same dispersion and convolved with a Gaussian to ensure that they
had the same spectral resolution as the science spectra. Final values
of mean velocity, velocity dispersion, $h_3$ and $h_4$ parameters were
assumed to be the average of the values for the three templates that
provided the closest fit (the lowest $\chi^2$ values).

In the pPXF analysis, two different sets of spectra were employed as
templates to extract the kinematical information, the aforementioned
Vazdekis et al. models, but this time the entire library consisting of
350 models, and the full MILES spectral library \citep{sanchez06}.
Both have a spectral resolution of $2.3$\,\AA\,FWHM \citep[but
see][]{beifiori11}. Since we expect to find a population gradient
across the galaxy (see Section \ref{results}), the templates used for
each radial bin were not fixed, but freely chosen by pPXF from the
MILES library and the Vazdekis models, separately, until the closest
model was achieved.  For the Vazdekis models, pPXF used typically 3
from the 350 models, while a mean number of 16 template stars from the
985 in the MILES library were used by pPXF for each radial
bin. Estimations of the errors for all the calculated values were
obtained by performing 100 Monte Carlo simulations in which random
noise was added to each science spectrum and then passed again through
pPXF \citep[e.g.][]{cappellari04, bedregal06}. In this case, only the
templates that were previously selected by pPXF were fitted, and not
the entire set. The dispersion in the values obtained from the
simulations are quoted as the error bars in the measurements from the
original spectra.

\begin{figure}[tp]
  \centering
  \includegraphics[width=0.50\textwidth]{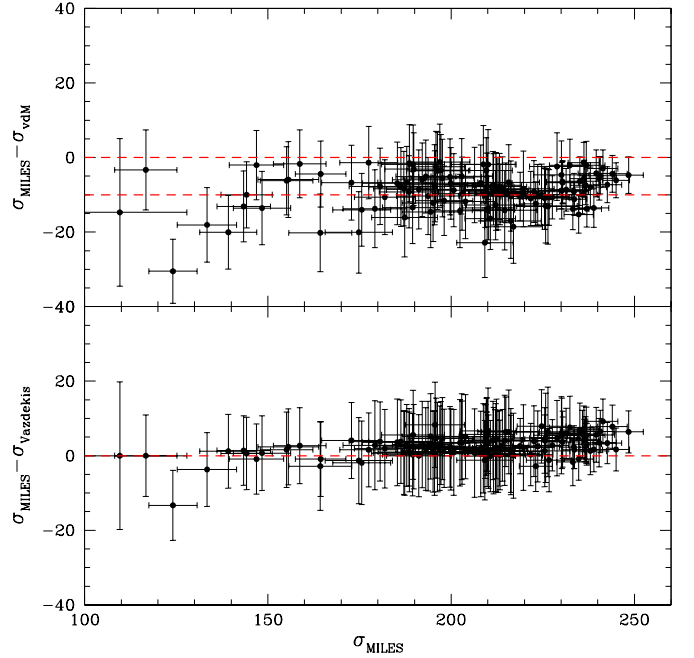}
  \caption{Comparison between the different measurements of the
    velocity dispersion profile of the major axis of NGC 7507 as
    described in the text.}
  \label{comparison}
\end{figure}

The agreement between the results of pPXF using the two sets of
templates is very good. However a few remarks are
appropriate. Typically pPXF uses much fewer templates from the
Vazdekis models ($\sim$3) and the $\chi^2$ values are always larger
than when using MILES. The inclusion of multiplicative Legendre
polynomials reduces the $\chi^2$, but does not increase the number of
templates used. We find a systematic small difference of $3$ km
s$^{-1}$ between the velocity dispersion results obtained using the
Vazdekis SSP models and the MILES library (Fig.~\ref{comparison},
bottom panel). The difference from the values given by the vdM code is
somewhat larger: the velocity dispersions derived with vdM are
systematically larger. We find a systematic difference of 9 km
s$^{-1}$ (Fig.~\ref{comparison}, top panel). Since template
mismatching is expected to be far less significant for this method, if
not totally negligible, based on previous studies
\citep[e.g.][]{shapiro06,cappellari07}, we decided to adopt the
results obtained using the MILES spectral in all the subsequent
kinematic description and dynamical analyses. These measurements can
be seen in Tables \ref{table:major} and \ref{table:minor}.

%%%%%%%%%%%%%%%%%%%%%%%%%%%%%%%%%%%%%%%%%%%%%%%%%%%%%%%%%%%%%%%%%%%%%%
\section{Results}
\label{results}
\subsection{Light and color profile }
\label{sec:light}

\begin{figure}[t!]
  \centering
   \includegraphics[width=0.50\textwidth]{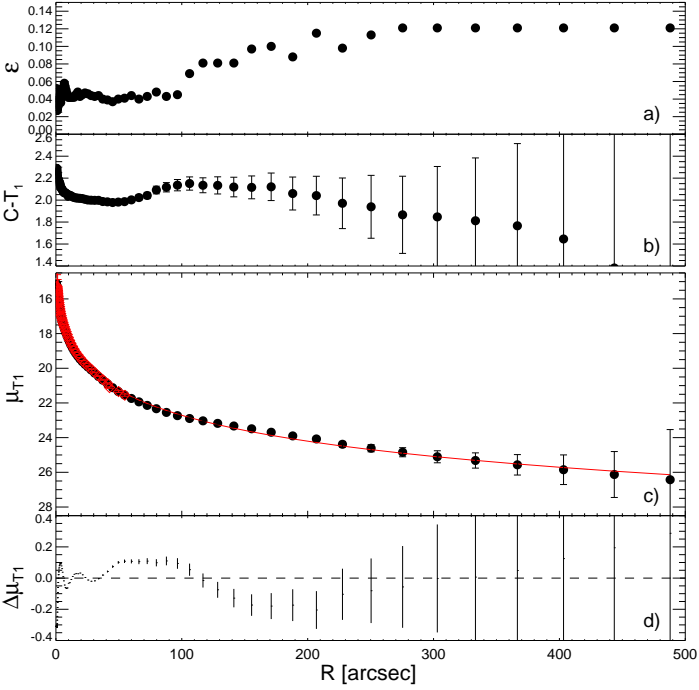}
   \caption{CTIO/MOSAIC surface photometry of NGC 7507. a) Ellipticity
     profile. b) $C-R$ color profile. c) $R$ surface brightness
     profile (black circles) and $R$ surface photometry from
     \citet{franx89a} (filled red squares). The solid red line is the
     fit of a double $\beta$ model to our data. Residuals can be seen
     in panel d.}
  \label{fig:photometry}
\end{figure}

There is a well-established variety of analytical model light profiles
of elliptical galaxies. Since we wish to use the light profile within
a spherical Jeans analysis, its analytical form, besides being an
excellent representation, should permit a closed analytical
deprojection as well as an analytical mass profile. These conditions
are fullfilled by the sum of two $\beta$ profiles (or Hubble-Reynolds
profile)
\begin{equation}\label{eq:light1}
  \mu(R)=-2.5\log\left\lbrace a_1\left(1+\left(\frac{R}{R_1}\right)^2\right)^{-1} + a_2\left(1+\left(\frac{R}{R_2}\right)^2\right)^{-1}\right\rbrace,
\end{equation}
where $a_1$\,=\,2.052$\times$10$^{-8}$,
$a_2$\,=\,6.452$\times$10$^{-7}$, $R_1$\,=\,16\arcsec, and
$R_2$\,=\,2.19\arcsec, provide an excellent representation of the
surface brightness profile between 2\arcsec and 400\arcsec
(Fig.~\ref{fig:photometry}, panel c).

Although the $\beta$-profile provides an excellent description out to
about 50 kpc, the extrapolation to larger radii, which is used in the
dynamic modeling, is not a priori justified.  It is generally unknown
where a galaxy ``ends'' (but see the remarks in
Sect. \ref{sec:stellar}).  The mass of the $\beta$-model diverges for
large radii. This raises the concern of whether systematic errors are
introduced by extrapolating the profile to larger radii.

We therefore fit the surface brightness profile with the sum of two
Sersic profiles
\begin{equation}\label{eq:sersic}
  I(R)=I_0\exp\left(-(R/a_s)^{1/m}\right)
\end{equation}
\citep{sersic68,graham05}, since a single Sersic profile fails to
accurately reproduce the luminosity in the entire radial range,
underestimating the outer ($\gtrsim$ 20 kpc) profile. The Sersic
parameters are $I_0^i=1.90\times10^6$ L$_{\odot}$ pc$^{-2}$,
$a_s^i=0.0677$ pc, $m^i=4.8$, for the inner Sersic profile, and
$I_0^o=15.77$ L$_{\odot}$ pc$^{-2}$, $a_s^o=15832$ pc, $m^o=1.05$, for
the outer profile; $M_{\sun,R}$\,=\,4.42 \citep{bm98} was adopted.

\begin{figure}[t]
  \centering
  \includegraphics[width=0.50\textwidth]{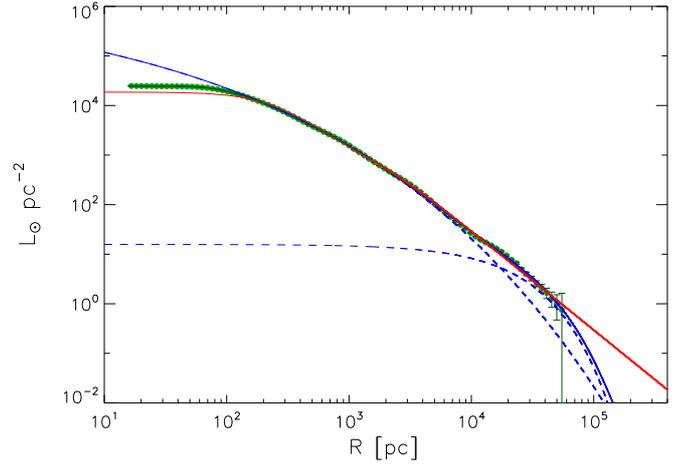}
  \caption{Surface brightness profile of NGC 7507. The green dots
    indicate the $R$ photometry in units of L$_{\odot}$ pc$^{-2}$. The
    dashed blue lines represent the two Sersic profiles, while the
    solid blue line is their sum. The red solid line represents the
    double $\beta$ model.}
  \label{fig:photometry2}
\end{figure}

Both models provide excellent descriptions of the surface density
profile from $\sim$ 200 pc ($\sim$1.5\arcsec) until the last
photometric point at $\sim$ 55 kpc, and are practically
indistinguishable (see Fig. \ref{fig:photometry2}). Differences are
seen only in the inner $\sim$200 pc where the double $\beta$ model
underestimates the light profile, while the opposite happens with the
double Sersic model and for the extrapolation of the light profile
(outside $\sim$ 55 kpc) where the double $\beta$ model falls as
$R^{-2.0}$, but the Sersic profile declines more steeply.

The double Sersic profile implies that the effective radius is
$R_e$=75\arcsec. This is significantly larger than the previous
estimations of 24\arcsec \citep{lauberts89}, 31\arcsec
\citep{faber89}, and 47\arcsec \citep{desouza04}; this is unsurprising
given our deep imaging. For example, the value of Faber et al. is
taken from the RC2 \citep{devaucouleurs76} where the total galaxy
brightness is defined as the brightness within an isophote with a
surface brightness in the $B$-band of 25 mag arcsec$^{-2}$ which we
found for NGC 7507 occurs at a distance of about 150\arcsec.

The color profile was evaluated as the difference between the two
models in $C$ and $R$ (Fig.~\ref{fig:photometry}, panel b). The inner
core (within 10$\arcsec$) declines by 0.3 mag with radius to bluer
colors, followed by a radial interval of more or less constant
color. The color then gets redder again at about 50$\arcsec$, and
remains practically unchanged out to 200$\arcsec$. The outer gradient
to bluer colors might exist, but owing to the large uncertainties
cannot be claimed convincingly. Color profiles for the inner
50$\arcsec$ were given by \citet{franx89a} and
\citet{goudfrooij94}. The inner strong gradient is apparent in both of
them, but the gradient of almost 0.5 mag in $B-I$ between 30$\arcsec$
and 60$\arcsec$ seen by \citet{goudfrooij94} is not found in our color
profile.

\begin{figure}[t!]
  \centering
  \includegraphics[width=0.475\textwidth]{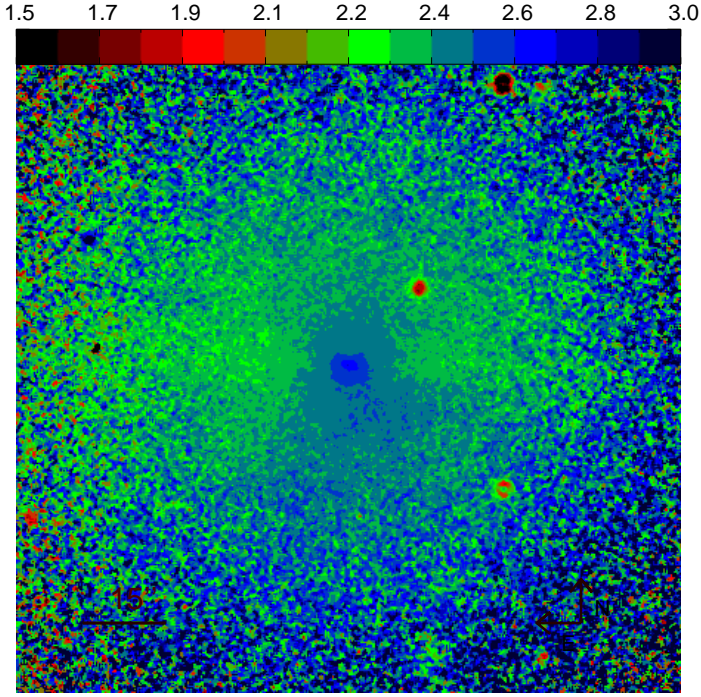}
  \caption{$C-R$ color image of NGC 7507 based on the short
    exposures. Image has been median smoothed in $3\times3$ pixel
    boxes. Image size is $2\arcmin\times2\arcmin$.}
  \label{fig:colorimage}
\end{figure}

It is interesting to consider a 2D color image of the inner region.
Fig.~\ref{fig:colorimage} was obtained by dividing the short $C$ by
the $R$ exposures. It is apparent that the ``color field'' is not
spherically symmetric but tends to have bluer colors along the E-W
axis. One could therefore interpret the plateau with $C-R$=2.0 between
10$\arcsec$ and 50$\arcsec$ as a result of averaging along isophotes,
before the color gets redder again. We are unable to determine,
however, whether diffuse dust or population effects or both are
responsible.

\subsection{Ellipticity and higher isophote moments}
\label{sec:ellip}

 \begin{figure}
  \centering
  \includegraphics[width=0.50\textwidth]{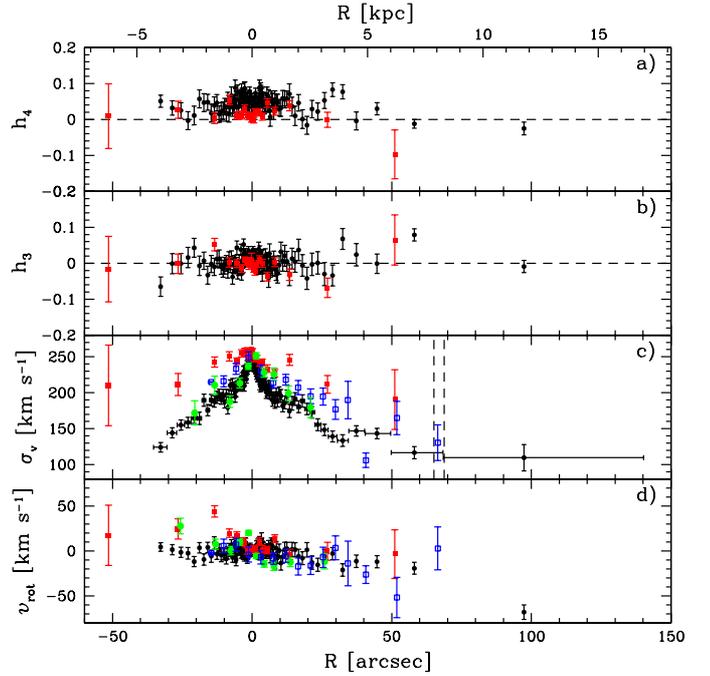}
  \caption{Major axis kinematics of NGC 7507, as extracted with pPXF
    (see text). a) Hermite coefficient $h_4$ profile, b) $h_3$
    profile, both panels show a dashed line at the zero level for
    comparison, c) velocity dispersion. The horizontal error bars show
    the regions across which the spatial binning was done (shown only
    in this panel, but valid for all four). The vertical dashed strip
    indicates a gap between the detectors in which no measurements are
    possible, d) rotation profile. For all panels, open black circles
    are our data, red solid squares are from
    \citetalias{kronawitter00}, solid green circles are from
    \citet{franx89b}, and blue open squares come from
    \citet{bertin94}.}
  \label{fig:major}
\end{figure}

As all previous studies have pointed out, the galaxy in projection is
very round. We find $\langle\epsilon \rangle=0.045\pm0.008$, where the
mean is taken over the inner 2\arcmin, avoiding the innermost
2$\arcsec$ which are affected by seeing. Beyond 2$\arcmin$ the
ellipticity slightly increases, partly due to the difficulty in
masking a bright $11^{\mathrm{th}}$ mag star, located $\sim$$6\arcmin$
SE of the galaxy (Fig~\ref{fig:photometry}, panel a).

The Fourier coefficient $a_4$ describes the isophote's boxiness or
diskyness (negative: boxy; positive:disky). It is zero out to
2$\arcmin$.  The slight boxiness at larger radii may also be caused by
the bright star. A table with the photometry, including ellipticities,
PA, and $a_4$ is available online.

\subsection{Irregular features}

A small, slightly off-centered dip in the light can be seen in the
shallow $R$ image. It is not apparent in the shallow $C$ image, which
has poorer seeing. We interpret this as a dust lane already detected
by \citet{franx89a}.

\citet{tal09} found evidence of a faint shell north of the
galaxy. However, when the smoothed 2D model created with
\verb+IRAF\BMODEL+ task is subtracted from the galaxy light, we see
thi shell in neither the $C$ nor in the $R$ images. No substructure
was detected either in the shallower images analyzed by
\citet{desouza04}.

\subsection{Stellar kinematics}
\label{sec:kinematics}
   
The kinematics along the major axis of NGC 7507 have been measured by
several groups \citep[][K+00]{bertin94,franx89b}, but along the minor
axis only by \citet{franx89b} and higher order moments only by
\citetalias{kronawitter00}. We extended the radial coverage of
\citet{franx89b} by a factor of 1.5 for the major axis and 2.5 for the
minor axis.

\begin{figure}
  \centering
  \includegraphics[width=0.50\textwidth]{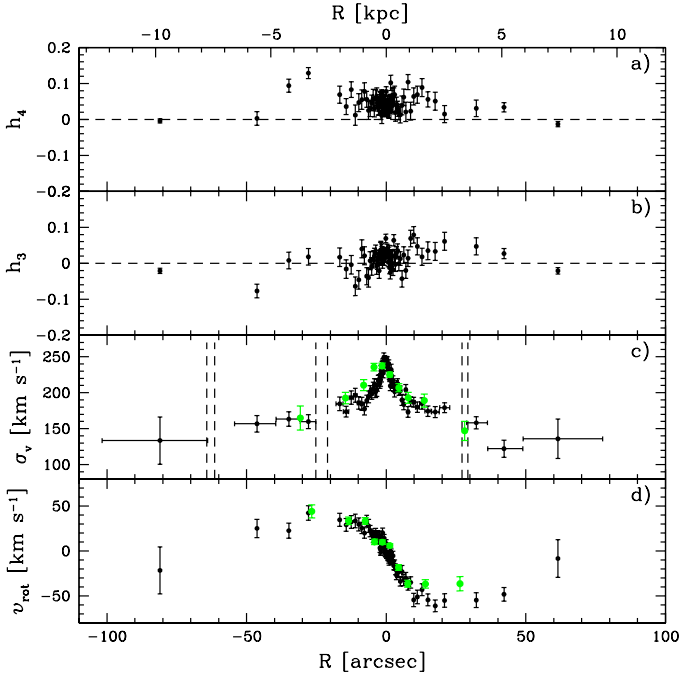}
  \caption{Minor axis kinematics of NGC 7507. The vertical dashed
    strips show the positions of the gaps between detectors and two
    interloping background galaxies, which were masked in the spatial
    binning process. Symbols are the same as in Fig. \ref{fig:major}}
  \label{fig:minor}
 \end{figure}

\subsubsection{Rotation}
\label{sec:rotation}

The major axis rotation is very low (Fig. \ref{fig:major}, panel d),
consistent with no rotation, with a formal amplitude of $-1 \pm 6$ km
s$^{-1}$ within the inner 30\arcsec. In the 30--60\arcsec\, range, the
amplitude slightly increases to a value of $-13 \pm 7$ km s$^{-1}$,
which is consistent with the aforementioned previous studies. The
outermost measured bin, at a mean distance of 97\arcsec shows a
significant increase in the rotation with respect to the center ($-68
$ $\pm$ 8 km s$^{-1}$). This may be partly due to the misalignment of
the slit direction and the photometric major axis but perhaps also a
hint that the kinematic properties of the galaxy suffer a change at
large radii, as seen for example in NGC 821 and NGC 1400
\citep{proctor09}.

The minor axis rotation is pronounced and slightly asymmetric with
amplitudes of $30 \pm 7$ km s$^{-1}$ in the west direction and $-43
\pm 7$ km s$^{-1}$ in the east direction inside the range 5-50\arcsec
(Fig. \ref{fig:minor}, panel d). \citet{franx89b} measured an
amplitude of 36 $\pm$ 5 km s$^{-1}$ in the range 5-30\arcsec, without
detecting any asymmetry. In the outermost east and west bins, around
60\arcsec, the mean velocity is consistent with no rotation. This
rotation profile closely resembles the major axis rotation of NGC 3379
\citep[][their Fig. 4]{statler99,weijmans09}.

Summarising, the isophote shapes are sufficiently undisturbed and the
projected rotation amplitude is small enough to justify the use of a
non-rotating, spherical mass model.
 
\subsubsection{Velocity dispersion}
\label{sec:vd}

The major axis velocity dispersion rapidly declines in the inner
30\arcsec and undergoes a slower decline at larger radii.  The
dispersion values of \citetalias{kronawitter00} deviate strongly, not
only from our values, but also from \citet{franx89b} and
\citet{bertin94} (Fig. \ref{fig:major}, panel c). While the difference
at the center of the galaxy is roughly within errors (9.7 $\pm$ 4.8 km
s$^{-1}$), and it can be easily explained by the spectra being taken
under different seeing conditions; at 27\arcsec\, from the center, the
difference amounts to 70 $\pm$ 14 km s$^{-1}$ in the sense that
\citetalias{kronawitter00} values are larger than ours.

The minor axis velocity dispersion profile shows a similar rapid
decline in the inner $30\arcsec$ but a less steep decline at larger
radii (Fig. \ref{fig:minor}, panel c). As in the case of the major
axis, the difference from \citet{bertin94} is minimal and amounts to
$14\pm 24$ km s$^{-1}$, where the uncertainty is caused almost
entirely by the \citet{bertin94} values.

A kinematic classification of early-type galaxies, based on an angular
momentum proxy was introduced by \citet{emsellem07}. They defined
\begin{equation}
 \lambda_R=\frac{\sum F_i R_i \left|V_i\right|}{\sum F_i R_i
  \sqrt{V_i^2 + \sigma_i^2}}, 
\end{equation} 
where $F_i$ is flux, $R_i$ is distance to the center of the galaxy,
and $V_i$ and $\sigma_i$ are the mean velocity and velocity
dispersion.  \citet{emsellem07} found that early-type galaxies can be
separated into two classes of slow and fast rotators, depending on
whether they have $\lambda_R$ values below (slow) or above (fast) 0.1
inside $R_e$. Even though this parameter was introduced for the SAURON
2D data, it can been applied to long-slit measurements in order to
help us to compare with the original definition, once a proper scaling
factor of 0.57 is applied \citep{cappellari07,coccato09}. In that case
the sum is over the quantities along the slit.

We calculated the $\lambda_R$ parameter separately for each long slit
position. For both axes, the parameter mildly increases inside $R_e$
reaching $\lambda_{R_e}^{\mathrm{min}}$\,=\,0.088 and
$\lambda_{R_e}^{\mathrm{maj}}$\,=\,0.018, staying basically flat
beyond that radius. This is consistent with the mild rotation of the
galaxy. NGC 7507 can be classified as a slow rotator, sharing also the
kinematic misalignment which is often seen in these galaxies, although
it has a rapidly declining velocity dispersion profile which is more
common amongst fast rotators \citep{coccato09}.

\subsubsection{Higher order moments}
\label{sec:higher}

The Gauss-Hermite coefficient $h_3$, which measures the asymmetric
deviation from a Gaussian, is consistent with being zero along the
major axis. Along the minor axis within $20\arcsec$, it appears to
have a gradient which would be consistent with its rotation
(Fig. \ref{fig:minor}, panels b and d).

The coefficient $h_4$, which indicates symmetric deviations from a
Gaussian profile, is positive for both the major and minor axis
spectra. The sign agrees with the measurements of
\citetalias{kronawitter00}, who also found slightly positive values of
$h_4$, although their values were somewhat smaller
(Fig. \ref{fig:major}, panel a).  We return to this result when
discussing any possible anisotropies.

\onltab{1}{
\begin{table*}
\caption{$T1$ CTIO/MOSAIC surface photometry of NGC 7507. First column indicates the semi-major axis from the ellipse fitting.}
\label{table:photometry}
\centering
\begin{tabular}{rccrr}
\hline\hline
\noalign{\smallskip} 
$R$& $\mu_{T1}$&$\epsilon$&$a_4$\,\,\,\,\,\,\,\,\,\,\,\,\,\,&PA\,\,\,\,\,\,\,\,\\
(arcsec)& (mag arcsec$^{-2}$)&&&\\
\hline
\noalign{\smallskip} 
0.148 & 15.136 $\pm$ 0.001 & 0.264 & 0.104 $\pm$ 0.043 & -23.9 $\pm$ 8.4\\
0.511 & 15.186 $\pm$ 0.001 & 0.052 & -0.023 $\pm$ 0.011 & -33.6 $\pm$ 16.3\\
1.095 & 15.550 $\pm$ 0.001 & 0.031 & -0.003 $\pm$ 0.001 & 8.3 $\pm$ 3.4\\
1.457 & 15.797 $\pm$ 0.001 & 0.033 & -0.001 $\pm$ 0.001 & 8.8 $\pm$ 2.1\\
2.133 & 16.194 $\pm$ 0.001 & 0.040 & -0.001 $\pm$ 0.001 & 13.8 $\pm$ 1.5\\
2.581 & 16.424 $\pm$ 0.001 & 0.043 & -0.001 $\pm$ 0.001 & 12.8 $\pm$ 1.3\\
3.123 & 16.678 $\pm$ 0.001 & 0.040 & -0.002 $\pm$ 0.001 & 15.4 $\pm$ 1.0\\
3.435 & 16.810 $\pm$ 0.001 & 0.039 & -0.001 $\pm$ 0.001 & 16.3 $\pm$ 1.2\\
4.572 & 17.200 $\pm$ 0.001 & 0.046 & -0.000 $\pm$ 0.001 & 14.4 $\pm$ 1.1\\
5.029 & 17.326 $\pm$ 0.001 & 0.048 & -0.000 $\pm$ 0.001 & 16.4 $\pm$ 0.8\\
5.532 & 17.449 $\pm$ 0.002 & 0.052 & -0.001 $\pm$ 0.001 & 15.1 $\pm$ 1.1\\
6.086 & 17.574 $\pm$ 0.002 & 0.054 & -0.002 $\pm$ 0.001 & 15.7 $\pm$ 0.9\\
7.364 & 17.837 $\pm$ 0.002 & 0.054 & 0.003 $\pm$ 0.001 & 15.6 $\pm$ 0.8\\
8.100 & 17.983 $\pm$ 0.003 & 0.050 & 0.001 $\pm$ 0.001 & 16.7 $\pm$ 0.8\\
8.910 & 18.138 $\pm$ 0.003 & 0.046 & 0.001 $\pm$ 0.001 & 14.7 $\pm$ 0.9\\
9.801 & 18.307 $\pm$ 0.004 & 0.042 & -0.001 $\pm$ 0.001 & 15.5 $\pm$ 0.9\\
10.782 & 18.475 $\pm$ 0.001 & 0.042 & 0.002 $\pm$ 0.001 & 12.1 $\pm$ 0.5\\
11.860 & 18.646 $\pm$ 0.001 & 0.042 & 0.002 $\pm$ 0.001 & 12.1 $\pm$ 0.5\\
13.046 & 18.817 $\pm$ 0.001 & 0.042 & 0.003 $\pm$ 0.001 & 12.1 $\pm$ 0.5\\
14.351 & 18.985 $\pm$ 0.001 & 0.042 & 0.003 $\pm$ 0.001 & 12.1 $\pm$ 0.6\\
15.786 & 19.145 $\pm$ 0.001 & 0.045 & 0.002 $\pm$ 0.001 & 13.0 $\pm$ 0.5\\
17.364 & 19.302 $\pm$ 0.001 & 0.048 & 0.001 $\pm$ 0.001 & 11.8 $\pm$ 0.5\\
19.101 & 19.468 $\pm$ 0.002 & 0.043 & 0.004 $\pm$ 0.001 & 10.3 $\pm$ 0.6\\
21.011 & 19.621 $\pm$ 0.002 & 0.045 & 0.003 $\pm$ 0.001 & 10.3 $\pm$ 0.5\\
23.112 & 19.772 $\pm$ 0.002 & 0.047 & 0.003 $\pm$ 0.001 & 8.4 $\pm$ 0.5\\
25.423 & 19.928 $\pm$ 0.003 & 0.046 & 0.002 $\pm$ 0.001 & 9.2 $\pm$ 0.5\\
27.965 & 20.096 $\pm$ 0.003 & 0.044 & 0.001 $\pm$ 0.001 & 10.3 $\pm$ 0.4\\
30.762 & 20.278 $\pm$ 0.003 & 0.043 & -0.001 $\pm$ 0.001 & 12.0 $\pm$ 0.4\\
33.838 & 20.473 $\pm$ 0.004 & 0.044 & 0.000 $\pm$ 0.001 & 11.9 $\pm$ 0.4\\
37.221 & 20.682 $\pm$ 0.005 & 0.040 & -0.000 $\pm$ 0.001 & 14.7 $\pm$ 0.5\\
40.944 & 20.894 $\pm$ 0.006 & 0.039 & -0.001 $\pm$ 0.001 & 16.5 $\pm$ 0.6\\
45.038 & 21.116 $\pm$ 0.008 & 0.037 & -0.002 $\pm$ 0.001 & 16.6 $\pm$ 0.7\\
49.542 & 21.336 $\pm$ 0.009 & 0.040 & -0.001 $\pm$ 0.001 & 17.2 $\pm$ 0.7\\
54.496 & 21.539 $\pm$ 0.011 & 0.041 & -0.001 $\pm$ 0.001 & 18.4 $\pm$ 0.8\\
59.946 & 21.736 $\pm$ 0.013 & 0.044 & -0.002 $\pm$ 0.001 & 19.9 $\pm$ 0.9\\
65.940 & 21.934 $\pm$ 0.016 & 0.040 & 0.000 $\pm$ 0.001 & 26.4 $\pm$ 1.0\\
72.534 & 22.138 $\pm$ 0.019 & 0.043 & -0.002 $\pm$ 0.001 & 23.1 $\pm$ 1.1\\
79.788 & 22.328 $\pm$ 0.023 & 0.048 & -0.004 $\pm$ 0.001 & 23.8 $\pm$ 1.1\\
87.766 & 22.541 $\pm$ 0.028 & 0.043 & -0.003 $\pm$ 0.001 & 25.3 $\pm$ 1.5\\
96.543 & 22.729 $\pm$ 0.034 & 0.045 & -0.005 $\pm$ 0.001 & 23.9 $\pm$ 1.7\\
106.197 & 22.894 $\pm$ 0.039 & 0.069 & -0.002 $\pm$ 0.002 & 28.2 $\pm$ 1.5\\
116.817 & 23.031 $\pm$ 0.045 & 0.081 & 0.002 $\pm$ 0.002 & 30.4 $\pm$ 1.3\\
128.499 & 23.177 $\pm$ 0.051 & 0.081 & 0.009 $\pm$ 0.002 & 37.7 $\pm$ 1.3\\
141.349 & 23.329 $\pm$ 0.059 & 0.081 & 0.000 $\pm$ 0.002 & 35.4 $\pm$ 1.3\\
155.484 & 23.490 $\pm$ 0.069 & 0.097 & -0.010 $\pm$ 0.002 & 28.5 $\pm$ 1.1\\
171.032 & 23.689 $\pm$ 0.083 & 0.100 & -0.000 $\pm$ 0.002 & 26.0 $\pm$ 1.3\\
188.135 & 23.900 $\pm$ 0.102 & 0.088 & -0.019 $\pm$ 0.003 & 28.4 $\pm$ 1.7\\
206.949 & 24.076 $\pm$ 0.121 & 0.115 & -0.028 $\pm$ 0.003 & 29.9 $\pm$ 1.1\\
227.643 & 24.381 $\pm$ 0.164 & 0.098 & -0.021 $\pm$ 0.003 & 34.7 $\pm$ 1.7\\
250.408 & 24.611 $\pm$ 0.207 & 0.113 & -0.020 $\pm$ 0.004 & 38.4 $\pm$ 1.7\\
275.449 & 24.842 $\pm$ 0.262 & 0.121 & 0.002 $\pm$ 0.004 & 36.4 $\pm$ 1.9\\
302.993 & 25.102 $\pm$ 0.346 & 0.121 & -0.087 $\pm$ 0.015 & 45.0 $\pm$ 3.9\\
333.293 & 25.318 $\pm$ 0.439 & 0.121 & -0.029 $\pm$ 0.007 & 38.1 $\pm$ 2.6\\
366.622 & 25.566 $\pm$ 0.589 & 0.121 & -0.001 $\pm$ 0.008 & 38.1 $\pm$ 3.6\\
403.284 & 25.849 $\pm$ 0.854 & 0.121 & 0.009 $\pm$ 0.009 & 38.1 $\pm$ 3.9\\
443.613 & 26.128 $\pm$ 1.324 & 0.121 & 0.027 $\pm$ 0.011 & 38.1 $\pm$ 4.3\\
487.974 & 26.428 $\pm$ 2.893 & 0.121 & 0.069 $\pm$ 0.056 & 38.1 $\pm$ 4.8\\
\hline
\end{tabular}
\end{table*}
}

\onltab{2}{
\begin{table*}
\caption{Major axis kinematics of NGC 7507.}
\label{table:major}
\centering
\begin{tabular}{rcccc}
\hline\hline
\noalign{\smallskip} 
$R$& $\varv_{\mathrm{rot}}$ & $\sigma_V$ & $h_3$ & $h_4$\\
(arcsec)& (km s$^{-1}$) & (km s$^{-1}$)& & \\
\hline
\noalign{\smallskip} 
-32.78 & 4.4 $\pm$ 4.8 & 124.1 $\pm$ 6.6 & -0.065 $\pm$ 0.027 & 0.051 $\pm$ 0.017\\
-28.68 & 1.5 $\pm$ 5.7 & 144.2 $\pm$ 7.0 & -0.001 $\pm$ 0.028 & 0.032 $\pm$ 0.020\\
-25.51 & -1.5 $\pm$ 5.7 & 155.1 $\pm$ 7.2 & -0.001 $\pm$ 0.028 & 0.025 $\pm$ 0.022\\
-22.96 & -2.5 $\pm$ 6.4 & 158.7 $\pm$ 7.2 & 0.016 $\pm$ 0.028 & -0.003 $\pm$ 0.025\\
-20.78 & -11.8 $\pm$ 5.7 & 164.4 $\pm$ 6.9 & 0.043 $\pm$ 0.026 & 0.011 $\pm$ 0.026\\
-18.88 & 3.5 $\pm$ 5.8 & 164.3 $\pm$ 8.6 & -0.007 $\pm$ 0.027 & 0.057 $\pm$ 0.025\\
-17.28 & -9.4 $\pm$ 6.3 & 189.6 $\pm$ 8.6 & 0.007 $\pm$ 0.024 & 0.047 $\pm$ 0.025\\
-15.90 & 4 $\pm$ 5.8 & 175.6 $\pm$ 7.9 & -0.033 $\pm$ 0.025 & 0.048 $\pm$ 0.022\\
-14.66 & -5.8 $\pm$ 6.7 & 189.8 $\pm$ 7.9 & -0.002 $\pm$ 0.024 & 0.018 $\pm$ 0.025\\
-13.57 & 9 $\pm$ 6.9 & 187.9 $\pm$ 7.9 & -0.010 $\pm$ 0.024 & 0.039 $\pm$ 0.024\\
-12.62 & -3.3 $\pm$ 6.0 & 193.7 $\pm$ 8.6 & 0.012 $\pm$ 0.023 & 0.046 $\pm$ 0.023\\
-11.82 & 1.2 $\pm$ 6.5 & 196.9 $\pm$ 8.2 & 0.012 $\pm$ 0.025 & 0.015 $\pm$ 0.023\\
-11.09 & -4.4 $\pm$ 7.6 & 192.1 $\pm$ 8.1 & -0.007 $\pm$ 0.024 & 0.021 $\pm$ 0.026\\
-10.43 & -3.1 $\pm$ 6.9 & 196.3 $\pm$ 8.5 & -0.015 $\pm$ 0.024 & 0.025 $\pm$ 0.024\\
-9.85 & -3.4 $\pm$ 6.1 & 187.1 $\pm$ 8.2 & -0.007 $\pm$ 0.023 & 0.036 $\pm$ 0.023\\
-9.26 & -8.8 $\pm$ 5.8 & 197.2 $\pm$ 7.9 & -0.000 $\pm$ 0.022 & 0.018 $\pm$ 0.023\\
-8.75 & -2.1 $\pm$ 7.0 & 179.2 $\pm$ 8.7 & -0.015 $\pm$ 0.026 & 0.053 $\pm$ 0.023\\
-8.31 & 1.8 $\pm$ 6.7 & 195.9 $\pm$ 7.7 & 0.008 $\pm$ 0.024 & 0.015 $\pm$ 0.022\\
-7.88 & -0.1 $\pm$ 5.9 & 210.1 $\pm$ 7.6 & -0.004 $\pm$ 0.022 & 0.041 $\pm$ 0.024\\
-7.44 & 6.3 $\pm$ 6.4 & 195.6 $\pm$ 8.0 & 0.014 $\pm$ 0.021 & 0.066 $\pm$ 0.024\\
-7.00 & -7.2 $\pm$ 5.4 & 211.6 $\pm$ 8.1 & -0.038 $\pm$ 0.020 & 0.034 $\pm$ 0.025\\
-6.64 & 0.2 $\pm$ 7.0 & 206.4 $\pm$ 8.2 & -0.005 $\pm$ 0.023 & 0.043 $\pm$ 0.026\\
-6.05 & -6.9 $\pm$ 6.1 & 194.4 $\pm$ 7.8 & -0.006 $\pm$ 0.022 & 0.088 $\pm$ 0.022\\
-5.47 & 1.4 $\pm$ 6.4 & 209.6 $\pm$ 7.1 & 0.043 $\pm$ 0.019 & 0.034 $\pm$ 0.022\\
-5.18 & 5.7 $\pm$ 6.1 & 212.4 $\pm$ 7.7 & -0.025 $\pm$ 0.019 & 0.076 $\pm$ 0.020\\
-4.59 & 7.8 $\pm$ 5.5 & 210.2 $\pm$ 7.1 & 0.006 $\pm$ 0.018 & 0.053 $\pm$ 0.019\\
-4.01 & -3 $\pm$ 5.1 & 214.9 $\pm$ 6.7 & 0.038 $\pm$ 0.015 & 0.064 $\pm$ 0.019\\
-3.50 & -0.1 $\pm$ 6.5 & 212.8 $\pm$ 8.0 & -0.012 $\pm$ 0.017 & 0.065 $\pm$ 0.022\\
-3.06 & -0.2 $\pm$ 5.9 & 228.9 $\pm$ 7.5 & 0.052 $\pm$ 0.016 & 0.041 $\pm$ 0.021\\
-2.04 & -1.7 $\pm$ 4.8 & 218.8 $\pm$ 5.2 & 0.005 $\pm$ 0.014 & 0.062 $\pm$ 0.015\\
0.00 & -0.2 $\pm$ 3.5 & 248.4 $\pm$ 4.0 & 0.026 $\pm$ 0.009 & 0.055 $\pm$ 0.011\\
1.02 & 2 $\pm$ 3.8 & 233.2 $\pm$ 4.4 & -0.010 $\pm$ 0.012 & 0.051 $\pm$ 0.013\\
2.04 & 7.8 $\pm$ 5.3 & 226.7 $\pm$ 6.2 & -0.009 $\pm$ 0.017 & 0.037 $\pm$ 0.015\\
3.07 & -0.2 $\pm$ 6.2 & 226.4 $\pm$ 8.5 & 0.012 $\pm$ 0.022 & 0.080 $\pm$ 0.022\\
3.50 & 7.1 $\pm$ 7.5 & 215.7 $\pm$ 8.4 & 0.011 $\pm$ 0.025 & 0.065 $\pm$ 0.021\\
4.16 & 3.2 $\pm$ 6.5 & 210.7 $\pm$ 6.9 & -0.001 $\pm$ 0.022 & 0.056 $\pm$ 0.020\\
4.45 & 0.9 $\pm$ 6.0 & 211.4 $\pm$ 7.8 & -0.006 $\pm$ 0.023 & 0.036 $\pm$ 0.021\\
5.03 & -1.5 $\pm$ 7.0 & 200.7 $\pm$ 8.5 & 0.000 $\pm$ 0.025 & 0.047 $\pm$ 0.026\\
5.62 & 1.0 $\pm$ 6.8 & 203.0 $\pm$ 8.2 & 0.004 $\pm$ 0.025 & 0.045 $\pm$ 0.023\\
5.91 & 4.4 $\pm$ 7.4 & 208.1 $\pm$ 8.5 & -0.041 $\pm$ 0.026 & 0.060 $\pm$ 0.023\\
6.50 & -5.5 $\pm$ 7.8 & 208.8 $\pm$ 8.7 & -0.011 $\pm$ 0.028 & 0.007 $\pm$ 0.026\\
6.86 & 7.9 $\pm$ 6.7 & 198.1 $\pm$ 7.8 & -0.005 $\pm$ 0.024 & 0.057 $\pm$ 0.022\\
7.73 & -4.9 $\pm$ 6.7 & 193.1 $\pm$ 8.2 & 0.007 $\pm$ 0.027 & 0.023 $\pm$ 0.023\\
8.17 & -12.8 $\pm$ 7.2 & 187.3 $\pm$ 9.0 & -0.008 $\pm$ 0.028 & 0.055 $\pm$ 0.025\\
8.61 & 3.2 $\pm$ 7.3 & 197.6 $\pm$ 8.6 & -0.020 $\pm$ 0.028 & 0.049 $\pm$ 0.022\\
9.12 & -7.5 $\pm$ 6.6 & 188.5 $\pm$ 8.1 & -0.028 $\pm$ 0.027 & 0.047 $\pm$ 0.023\\
9.70 & 0.5 $\pm$ 7.2 & 199.7 $\pm$ 8.5 & -0.005 $\pm$ 0.026 & 0.033 $\pm$ 0.023\\
10.29 & -13.4 $\pm$ 7.9 & 195.5 $\pm$ 9.0 & 0.032 $\pm$ 0.027 & 0.032 $\pm$ 0.023\\
10.94 & -6.4 $\pm$ 7.0 & 181.9 $\pm$ 8.3 & 0.006 $\pm$ 0.028 & 0.059 $\pm$ 0.022\\
11.67 & 1.5 $\pm$ 8.2 & 186.3 $\pm$ 8.8 & 0.021 $\pm$ 0.029 & 0.056 $\pm$ 0.024\\
12.47 & 2.6 $\pm$ 7.2 & 195.1 $\pm$ 8.5 & 0.005 $\pm$ 0.028 & 0.058 $\pm$ 0.021\\
13.35 & -5.0 $\pm$ 7.6 & 174.8 $\pm$ 9.2 & -0.033 $\pm$ 0.029 & 0.071 $\pm$ 0.028\\
14.30 & 1.9 $\pm$ 7.9 & 185.3 $\pm$ 9.5 & 0.013 $\pm$ 0.029 & 0.036 $\pm$ 0.026\\
15.39 & 1.4 $\pm$ 7.8 & 191.2 $\pm$ 8.8 & 0.004 $\pm$ 0.029 & 0.010 $\pm$ 0.027\\
16.63 & -5.9 $\pm$ 7.8 & 180.6 $\pm$ 8.4 & 0.037 $\pm$ 0.030 & 0.046 $\pm$ 0.023\\
18.01 & -8.0 $\pm$ 7.2 & 188.6 $\pm$ 8.5 & -0.006 $\pm$ 0.030 & 0.001 $\pm$ 0.026\\
19.62 & 1.3 $\pm$ 8.0 & 177.5 $\pm$ 7.9 & -0.042 $\pm$ 0.031 & -0.016 $\pm$ 0.025\\
21.44 & -13 $\pm$ 7.4 & 172.8 $\pm$ 8.1 & -0.003 $\pm$ 0.031 & 0.029 $\pm$ 0.024\\
23.48 & -15.4 $\pm$ 7.3 & 155.6 $\pm$ 8.3 & 0.001 $\pm$ 0.031 & 0.022 $\pm$ 0.024\\
25.88 & -6.2 $\pm$ 7.8 & 148.4 $\pm$ 7.8 & -0.030 $\pm$ 0.032 & 0.053 $\pm$ 0.022\\
28.78 & -2.8 $\pm$ 6.8 & 139.2 $\pm$ 7.8 & -0.034 $\pm$ 0.029 & 0.083 $\pm$ 0.019\\
32.40 & -21.1 $\pm$ 6.9 & 133.4 $\pm$ 8.1 & 0.068 $\pm$ 0.029 & 0.077 $\pm$ 0.020\\
37.30 & -11.4 $\pm$ 7.0 & 146.9 $\pm$ 7.4 & 0.024 $\pm$ 0.031 & -0.004 $\pm$ 0.025\\
44.70 & -12 $\pm$ 6.8 & 143.4 $\pm$ 7.3 & -0.001 $\pm$ 0.027 & 0.030 $\pm$ 0.017\\
58.08 & -19.3 $\pm$ 6.6 & 116.7 $\pm$ 8.5 & 0.079 $\pm$ 0.017 & -0.012 $\pm$ 0.012\\
97.33 & -68.0 $\pm$ 8.1 & 109.6 $\pm$ 18.3 & -0.009 $\pm$ 0.017 & -0.025 $\pm$ 0.018\\
\hline
\end{tabular}
\end{table*}
}

\onltab{3}{
\begin{table*}
\caption{Minor axis kinematics of NGC 7507.}
\label{table:minor}
\centering
\begin{tabular}{ccccc}
\hline\hline
\noalign{\smallskip} 
$R$& $\varv_{\mathrm{rot}}$ & $\sigma_V$ & $h_3$ & $h_4$\\
(arcsec)& (km s$^{-1}$) & (km s$^{-1}$)& & \\
\hline
\noalign{\smallskip} 
-81.02 & -21.64 $\pm$ 26.2 & 133.5 $\pm$ 32.8 & -0.021 $\pm$ 0.007 & -0.004 $\pm$ 0.006\\
-46.29 & 25.06 $\pm$ 10.2 & 156.8 $\pm$ 11.5 & -0.077 $\pm$ 0.019 & 0.003 $\pm$ 0.019\\
-34.90 & 22.56 $\pm$ 8.5 & 163.3 $\pm$ 10.1 & 0.008 $\pm$ 0.023 & 0.094 $\pm$ 0.018\\
-27.88 & 42.26 $\pm$ 8.0 & 159.9 $\pm$ 9.7 & 0.018 $\pm$ 0.023 & 0.129 $\pm$ 0.015\\
-16.66 & 34.56 $\pm$ 7.2 & 184.5 $\pm$ 9.5 & 0.017 $\pm$ 0.026 & 0.069 $\pm$ 0.024\\
-14.34 & 28.96 $\pm$ 7.2 & 173.6 $\pm$ 7.5 & -0.016 $\pm$ 0.026 & 0.036 $\pm$ 0.022\\
-12.53 & 32.26 $\pm$ 7.1 & 193.2 $\pm$ 9.4 & -0.004 $\pm$ 0.026 & 0.083 $\pm$ 0.022\\
-11.07 & 33.36 $\pm$ 7.5 & 196.5 $\pm$ 9.6 & -0.064 $\pm$ 0.026 & 0.012 $\pm$ 0.028\\
-9.84 & 29.56 $\pm$ 7.2 & 185.9 $\pm$ 8.6 & -0.046 $\pm$ 0.026 & 0.047 $\pm$ 0.025\\
-8.74 & 25.46 $\pm$ 6.3 & 184.5 $\pm$ 9.1 & 0.040 $\pm$ 0.025 & 0.055 $\pm$ 0.027\\
-7.80 & 19.86 $\pm$ 5.8 & 177.5 $\pm$ 8.4 & 0.021 $\pm$ 0.025 & 0.078 $\pm$ 0.024\\
-7.00 & 33.46 $\pm$ 6.1 & 188.8 $\pm$ 7.9 & -0.036 $\pm$ 0.024 & 0.056 $\pm$ 0.022\\
-6.35 & 27.86 $\pm$ 7.2 & 194.4 $\pm$ 8.3 & -0.040 $\pm$ 0.026 & 0.025 $\pm$ 0.025\\
-5.76 & 21.96 $\pm$ 6.3 & 201.5 $\pm$ 8.4 & 0.007 $\pm$ 0.025 & 0.031 $\pm$ 0.023\\
-5.18 & 18.76 $\pm$ 5.8 & 198.5 $\pm$ 7.5 & -0.009 $\pm$ 0.024 & 0.046 $\pm$ 0.022\\
-4.67 & 15.76 $\pm$ 6.4 & 205.7 $\pm$ 8.9 & 0.011 $\pm$ 0.024 & 0.058 $\pm$ 0.026\\
-4.23 & 16.76 $\pm$ 5.9 & 211.4 $\pm$ 8.1 & 0.020 $\pm$ 0.022 & 0.030 $\pm$ 0.023\\
-3.79 & 20.16 $\pm$ 5.0 & 201.2 $\pm$ 8.0 & 0.029 $\pm$ 0.022 & 0.060 $\pm$ 0.022\\
-3.43 & 14.56 $\pm$ 6.0 & 204.1 $\pm$ 9.2 & -0.007 $\pm$ 0.024 & 0.063 $\pm$ 0.025\\
-3.14 & 16.56 $\pm$ 6.1 & 210.7 $\pm$ 8.4 & 0.005 $\pm$ 0.022 & 0.046 $\pm$ 0.023\\
-2.04 & 1.26 $\pm$ 6.2 & 217.9 $\pm$ 8.8 & 0.035 $\pm$ 0.023 & 0.053 $\pm$ 0.022\\
-1.02 & 7.76 $\pm$ 4.8 & 227.4 $\pm$ 6.7 & 0.022 $\pm$ 0.017 & 0.064 $\pm$ 0.016\\
0.00 & 8.36 $\pm$ 4.0 & 245.4 $\pm$ 5.3 & 0.029 $\pm$ 0.012 & 0.078 $\pm$ 0.013\\
1.02 & -11.04 $\pm$ 4.9 & 224.5 $\pm$ 5.3 & 0.034 $\pm$ 0.016 & 0.023 $\pm$ 0.019\\
2.11 & -10.44 $\pm$ 4.3 & 212.1 $\pm$ 5.9 & 0.019 $\pm$ 0.015 & 0.067 $\pm$ 0.020\\
3.28 & -20.44 $\pm$ 6.0 & 207.2 $\pm$ 7.1 & -0.003 $\pm$ 0.020 & 0.019 $\pm$ 0.024\\
3.64 & -26.94 $\pm$ 5.1 & 209.8 $\pm$ 6.9 & 0.029 $\pm$ 0.017 & 0.042 $\pm$ 0.020\\
4.08 & -23.24 $\pm$ 6.4 & 212.7 $\pm$ 6.6 & 0.035 $\pm$ 0.019 & 0.030 $\pm$ 0.021\\
4.52 & -23.14 $\pm$ 5.9 & 203.1 $\pm$ 6.6 & -0.001 $\pm$ 0.020 & 0.021 $\pm$ 0.022\\
5.03 & -33.94 $\pm$ 5.2 & 202.8 $\pm$ 6.4 & 0.013 $\pm$ 0.019 & 0.013 $\pm$ 0.023\\
5.62 & -22.94 $\pm$ 6.9 & 189.9 $\pm$ 7.7 & -0.043 $\pm$ 0.023 & 0.039 $\pm$ 0.026\\
6.27 & -32.24 $\pm$ 6.5 & 183.9 $\pm$ 7.5 & 0.023 $\pm$ 0.023 & 0.062 $\pm$ 0.023\\
7.00 & -30.14 $\pm$ 6.2 & 204.0 $\pm$ 7.4 & -0.020 $\pm$ 0.022 & 0.021 $\pm$ 0.027\\
7.80 & -36.94 $\pm$ 6.6 & 173.3 $\pm$ 7.3 & 0.014 $\pm$ 0.023 & 0.104 $\pm$ 0.021\\
8.74 & -35.04 $\pm$ 6.2 & 187.7 $\pm$ 6.4 & 0.069 $\pm$ 0.023 & 0.023 $\pm$ 0.025\\
9.84 & -54.44 $\pm$ 7.4 & 186.8 $\pm$ 7.0 & 0.079 $\pm$ 0.023 & 0.064 $\pm$ 0.024\\
11.14 & -51.14 $\pm$ 7.6 & 179.9 $\pm$ 7.2 & 0.047 $\pm$ 0.024 & 0.069 $\pm$ 0.024\\
12.81 & -43.04 $\pm$ 6.6 & 184.9 $\pm$ 7.9 & 0.018 $\pm$ 0.024 & 0.089 $\pm$ 0.025\\
14.92 & -54.44 $\pm$ 6.6 & 174.4 $\pm$ 7.3 & 0.035 $\pm$ 0.025 & 0.056 $\pm$ 0.021\\
17.53 & -61.14 $\pm$ 6.5 & 173.2 $\pm$ 7.7 & 0.033 $\pm$ 0.025 & 0.051 $\pm$ 0.025\\
20.87 & -55.14 $\pm$ 7.5 & 179.3 $\pm$ 6.9 & 0.061 $\pm$ 0.025 & 0.015 $\pm$ 0.024\\
32.23 & -54.74 $\pm$ 8.3 & 158.2 $\pm$ 8.3 & 0.047 $\pm$ 0.024 & 0.031 $\pm$ 0.023\\
42.16 & -48.24 $\pm$ 7.5 & 122.1 $\pm$ 11.9 & 0.027 $\pm$ 0.014 & 0.034 $\pm$ 0.013\\
61.48 & -8.44 $\pm$ 21.1 & 135.9 $\pm$ 27.4 & -0.021 $\pm$ 0.009 & -0.013 $\pm$ 0.007\\
\hline
\end{tabular}
\end{table*}
}

%%%%%%%%%%%%%%%%%%%%%%%%%%%%%%%%%%%%%%%%%%%%%%%%%%%%%%%%%%%%%%%%%%%%%%
\section{Dynamical modeling}
\label{sec:models}

\subsection {The stellar mass profile}
\label{sec:stellar}

To obtain the (spherical) stellar mass profile we have to deproject
the surface brightness profile after correcting for Galactic
absorption \citep[$A_R=0.128$;][]{schlegel98}. In the case of the
double-$\beta$ profile this is achieved by applying the Abel
deprojection formula, which for the $\beta$ profile has an analytical
description
\begin{equation}\label{eq:light2}
j(r)=\frac{C_R}{2}\sum_{i=1}^2\frac{a_i/R_i}{[1+(r/R_i)^2]^{3/2}},
\end{equation} 
where $a_i$ and $R_i$ come from Eq. \ref{eq:light1}, and
$C_R$\,=\,$2.5\times10^{10}$ is the factor used to transform to units
of L$_{\sun}$\,pc$^{-2}$, adopting $M_{\sun,R}$\,=\,4.42
\citep{bm98}. The enclosed stellar mass can be obtained by integrating
this light density

\begin{equation}\label{eq:starmass1}
M_{\star}(r)=4\pi\int_o^r\Upsilon_{\star,R}(r)j(r)r^2\mathrm{d}r,
\end{equation}
where $\Upsilon_{\star,R}$ is the stellar $M/L_R$. For constant
$M/L_R$, this integral has the exact solution
\begin{equation}\label{eq:starmass2}
M_{\star}(r)=2\pi C_R\Upsilon_{\star,R}\sum_{i=1}^2 a_i R_i^2\left(\sinh^{-1}x_i-\frac{x_i}{\sqrt{x_i^2+1}}\right),
\end{equation}
where $x_i=r/R_i$ and $a_i$ and $R_i$ have the same meaning as in
Eq. \ref{eq:light2}. 

The stellar $M/L_R$ appears as a parameter in the fit to the
kinematical data, adopting the deprojected light profile and certain
forms for a possible DM halo.  From the shape of the colour
profile, one concludes that $M/L$ as a population property is
approximately constant out to a radius of 2$\arcmin$. At larger radii,
the $M/L$ may drop owing to a declining metallicity.  However, this is
only weakly constrained and any effect on the inner projected velocity
dispersion would be completely degenerate with a dark halo.

The Sersic profile can be deprojected using an approximation given
  by \citet{prugniel97} 
\begin{eqnarray*}
  j(r)&=&j_1\tilde{j}(r/a_s),\label{eq:sersic_dep}\\
  \tilde{j}(x)&\simeq& x^{-p}\mathrm{exp}(-x^{1/m}),\\
  j_1&=&\left\lbrace\frac{\Gamma(2m)}{\Gamma[(3-p)m]}\right\rbrace\frac{I_0}{2a_s
  },\\
 p&\simeq& 1 - 0.6097/m + 0.05463/m^2,
\end{eqnarray*}
where the latter equation comes from \citet{lima99}, $\Gamma(x)$ is
the Gamma function, and $I_0$, $a_s$, and $m$ refer to the Sersic
parameters from Eq. \ref{eq:sersic}. We also refer to \citet{mamon05b}
for a compilation of the relevant formulae.

Integrating the luminosity density, $j(r)$, the enclosed luminosity is
found.  \citet{lima99} give  
\begin{eqnarray}
  L_s(r)&=&L_{\mathrm{tot}}\tilde{L}_s(r/a_s),\label{eq:sersic_lum}\\
\tilde{L}_s(x)&=&\frac{\gamma[(3-p)m,x^{1/m}]}{\Gamma[(3-p)m]},
\end{eqnarray}
where $\gamma$ is now the incomplete Gamma function and
\begin{equation} \label{eq:total_sersic}
  L_{\mathrm{tot}}=2\pi m\Gamma(2m)I_0a_s^2 
\end{equation}
is the total luminosity of the profile. For our double Sersic profile,
the luminosity is simply the sum of the luminosity of both the inner
and outer Sersic profiles.

\begin{figure}[t!]
   \centering
\includegraphics[width=0.49\textwidth]{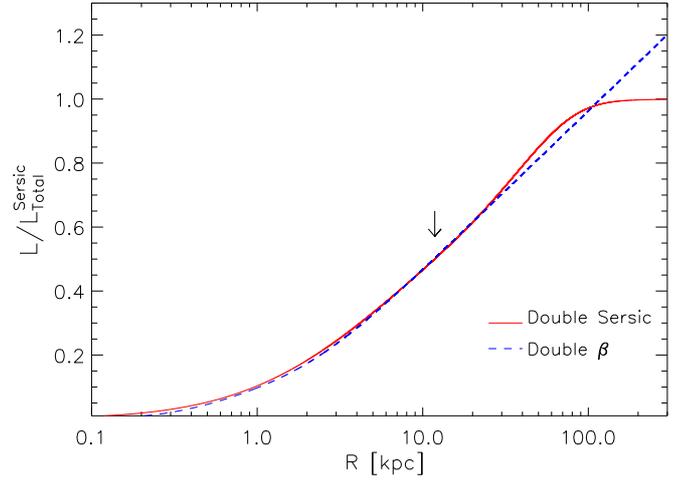}
\caption{A comparsion between the enclosed luminosities of the double
  $\beta$ profile (dashed blue line) and the double Sersic profile
  (red solid line), normalized to the double Sersic total
  luminosity. The arrow indicates the last spectroscopic point
  measured on the major axis.}
\label{fig:sersic_beta2}
    \end{figure}

Since beyond $\sim$55 kpc the double-$\beta$ and Sersic profiles
display the most distinct behaviors (cf. Fig. \ref{fig:photometry2}),
we inspected the influence of these dissimilar extrapolations on their
respective enclosed luminosities, which are directly related to the
stellar mass through the mass-to-light ratio. Fig.
\ref{fig:sersic_beta2} considers the enclosed luminosities of both
models normalized to the total luminosity of the double Sersic model
given by Eq. \ref{eq:total_sersic}. As expected from
Fig. \ref{fig:photometry2}, the luminosities are very similar not only
within the $\sim$55 kpc where the galaxy light was directly measured,
but well beyond. A $\sim$10\% difference in the total enclosed
luminosity is reached only at 200 kpc. In practice, however, the
extension of elliptical galaxies may be lower than this limit. In NGC
4636, for example, the globular cluster distribution is cut-off at
about $\sim$60 kpc \citep{dirsch05}. A similar behavior has been seen
in M 87, the central galaxy of Virgo, where the planetary nebulae
distribution show a truncation of the stellar halo at $\sim$150 kpc
\citep{doherty09}. Therefore, any differences in the extrapolations
beyond 200 kpc are not relevant to a dynamical analysis of the galaxy.

Given the similarities between the light profiles and to avoid any
uncertainties introduced by the approximations used in the Sersic
profile deprojection (which are not present in the analytic form of
the $\beta$ profile deprojection), the Jeans analysis that we now
describe was based on the stellar distribution given by the $\beta$
model profile. In any case, the velocity dispersions shown in
Fig. \ref{fig:dyn_models} were found to vary by less than 2 km
s$^{-1}$ at the most distant kinematic point, when the double Sersic
model was used a basis of the stellar mass profile.

\subsection{Jeans analysis}

The basis of our modeling is the spherical Jeans equation, which is
strictly valid only for infinite, non-rotating systems.  Its
solutions, when applied to real galaxies, may be considered as good
approximations, but there is no guarantee that they correspond to
physically correct distribution functions. Moreover, the potential and
the orbital anisotropy are degenerate, although we refer to
\citet{hansen06} and the remarks below. On the positive side, this
equation has an appealing simplicity, known anisotropies of elliptical
galaxies in the inner regions are generally small
\citep[e.g.][]{gerhard01,cappellari06} and one can use the higher
moments of the velocity distribution as constraints of the
anisotropy. \citet{cappellari06} also demonstrated that more
sophisticated dynamical modeling does not reveal grossly different
results. The main uncertainty probably does not arise from the
modeling approach, but collectively from the distance uncertainty,
the quality of the data, and the assumption of sphericity.
     
For convenience and clarity, we give in the following the basic
formulae.  The Jeans equation is \citep[e.g.][]{bt08}
\begin{equation}
\frac{\mathrm{d}[j(r)\sigma_r^2(r)]}{\mathrm{d} r} + \frac{2\beta(r)}{r}j(r)\sigma_r^2(r)=-j(r)\frac{GM(r)}{r^2},
\end{equation}
which relates the light distribution, $j(r)$, and radial velocity
dispersion, $\sigma_r(r)$, with the underlying gravitational potential
or enclosed total mass, $M(r)$, and where
\begin{equation}
\beta(r)=1-\sigma_{\theta}^2/\sigma_{r}^2
\end{equation}
is the anisotropy parameter, which indicates possible departures from
pure isotropy in the form of radial ($\beta\!>\!0$) or tangential
($\beta\!<\!0$) orbits.

For a given light and mass distribution, and a constant anisotropy,
the solution to the Jeans equation is
\begin{equation}\label{eq:jeans2}
j(r)\sigma_r^2(r)=G\int_r^{\infty}\frac{j(s)M(s)}{s^2}\left(\frac{s}{r}\right)^{2\beta}\mathrm{d}s
\end{equation}
\citep[e.g.][hereafter M\L05]{mamon05b}, which is then projected to be compared with the
velocity dispersion measurements
\begin{equation}\label{eq:jeans3}
\sigma_{\mathrm{LOS}}^2(R)=\frac{2}{I(R)}\left[\int_R^{\infty}\frac{j\sigma_r^2r\mathrm{d}r}{\sqrt{r^2-R^2}} - R^2\int_R^{\infty}\frac{\beta j\sigma_r^2\mathrm{d}r}{r\sqrt{r^2-R^2}}\right]
\end{equation}
\citep{binney82}. For the anisotropy profiles adopted in
Sects. \ref{sec:stars_only} and \ref{sec:dark_halos},
Eqs. \ref{eq:jeans2} and \ref{eq:jeans3} can be combined to yield the
single integral solution
\begin{equation}\label{eq:jeans4}
  \sigma_{\mathrm{LOS}}^2(R)=\frac{2}{I(R)}\int_R^{\infty}K(R,r)j(r)M(r)\frac{\mathrm{dr}}{r},
\end{equation}
where the kernels $K$ for the different anisotropy models are provided
in Appendix A2 of \citep{mamon05b}.

The accuracy of the models can be quantified by the use of a simple
$\chi^2$ test
\begin{equation}\label{eq:chi2}
  \chi^2=\sum_i^{N_{\mathrm{data}}}\left(\frac{\sigma_i^{\mathrm{obs}} - \sigma_i^{\mathrm{model}}}{\delta\sigma_i^{\mathrm{obs}}}\right)^2,
\end{equation}
where $\sigma^{\mathrm{obs}}$ is the measured velocity dispersion,
$\sigma^{\mathrm{model}}$ is the predicted dispersion at the same radial
distance, $\delta\sigma^{\mathrm{obs}}$ is the observational error associated
with the dispersion measurement, and the sum is taken over all  radial
bins.  

Under the spherical Jeans analysis, additional constraints on the
dynamical behavior of the galaxy can be obtained by modeling the
higher order moment of the velocity distribution which can be related
to the Gauss-Hermite coefficient $h_4$ \citep{vdm93}. For this, the
higher order Jeans equations must be solved. This task can be achieved
only if additional information about the distribution function is
known. In particular, by assuming distribution functions of the form
$f(E,L)=f_o(E)L^{-2\beta}$ with \emph{constant} $\beta$, it is
possible to reduce the two higher-order Jeans equation to only one. A
more detailed discussion, as well as the relevant formulae can be seen
in \citet{lokas02} and \citet{napolitano09}. Our results using this
approach are presented and discussed in Sect.~\ref{sec:h4}.

\subsection{The velocity dispersion profile}

\begin{table}
  \caption{Adopted velocity dispersion,$\sigma$, and $h_4$ values and their
    respective uncertainties. We refrain from
    giving the $h_3$-values since they are not modelled.}
  \label{tab:dispersions}
  \centering
  \begin{tabular}{lllrr}
    \hline\hline
    \noalign{\smallskip}
    Radius  &   $\sigma$ & $\Delta \sigma $ & $h_4$ & $\Delta h_4$\\ 
    (arcsec)& (km s$^{-1}$)&(km s$^{-1}$) & &\\
    \hline
    0.5 &    239 &  2 &    0.051 &  0.013 \\                      
    1.5 &    232 &  2 &    0.045 &  0.013 \\
    2.5 &    222 &  2 &    0.010 &  0.017 \\
    3.5 &    217 &  2 &    0.005 &  0.018 \\
    4.5 &    211 &  2 &    0.002 &  0.011 \\
    5.5 &    205 &  2 &    0.055 &  0.014 \\
    6.5 &    202 &  2 &    0.043 &  0.027 \\
    7.5 &    199 &  2 &    0.042 &  0.016 \\
    8.5 &    189 &  2 &    0.043 &  0.019 \\
    9.5 &    193 &  2 &    0.033 &  0.012 \\
    10  &    191 &  2 &    0.040 &  0.015 \\
    13  &    190 &  2 &    0.048 &  0.026 \\
    16  &    181 &  2 &    0.037 &  0.026 \\
    22  &    167 &  2 &    0.019 &  0.023 \\
    28  &    153 &  2   &  0.042 &  0.036 \\
    42  &    141 &  3.3 &  0.044 &  0.033 \\
    52  &    135 &  6.7 &  0.008 &  0.022 \\
    62  &    131 &  14.3& $-0.012$ &  0.036 \\
    85  &    125 &  18.7& $-0.013$ &  0.015 \\

    \noalign{\smallskip}
    \hline
  \end{tabular}

\end{table}

To construct the radial dependence of the velocity dispersion profile,
we averaged for a given radial bin the dispersion along the major and
the minor axis (although the ellipticity is tiny) from
Figs. \ref{fig:major} and \ref{fig:minor}. The dispersion values are
symmetrical with respect to the galaxy center and are in agreement
between the both axes when errors are considered. The bins are chosen
in such a way that the inner region, where the S/N is highest, has the
densest binning while the outer radii are covered with larger
bins. The principal uncertainties are those in the mean values
calculated by pPXF. Final values are listed in
Table~\ref{tab:dispersions}, which also lists the adopted $h_4$
values. Their uncertainties are the standard deviations of the $h_4$
values in the respective bins.

\subsection{Models without DM}
\label{sec:stars_only} 

\begin{figure*}[htp!]
\centering
\includegraphics[width=\textwidth]{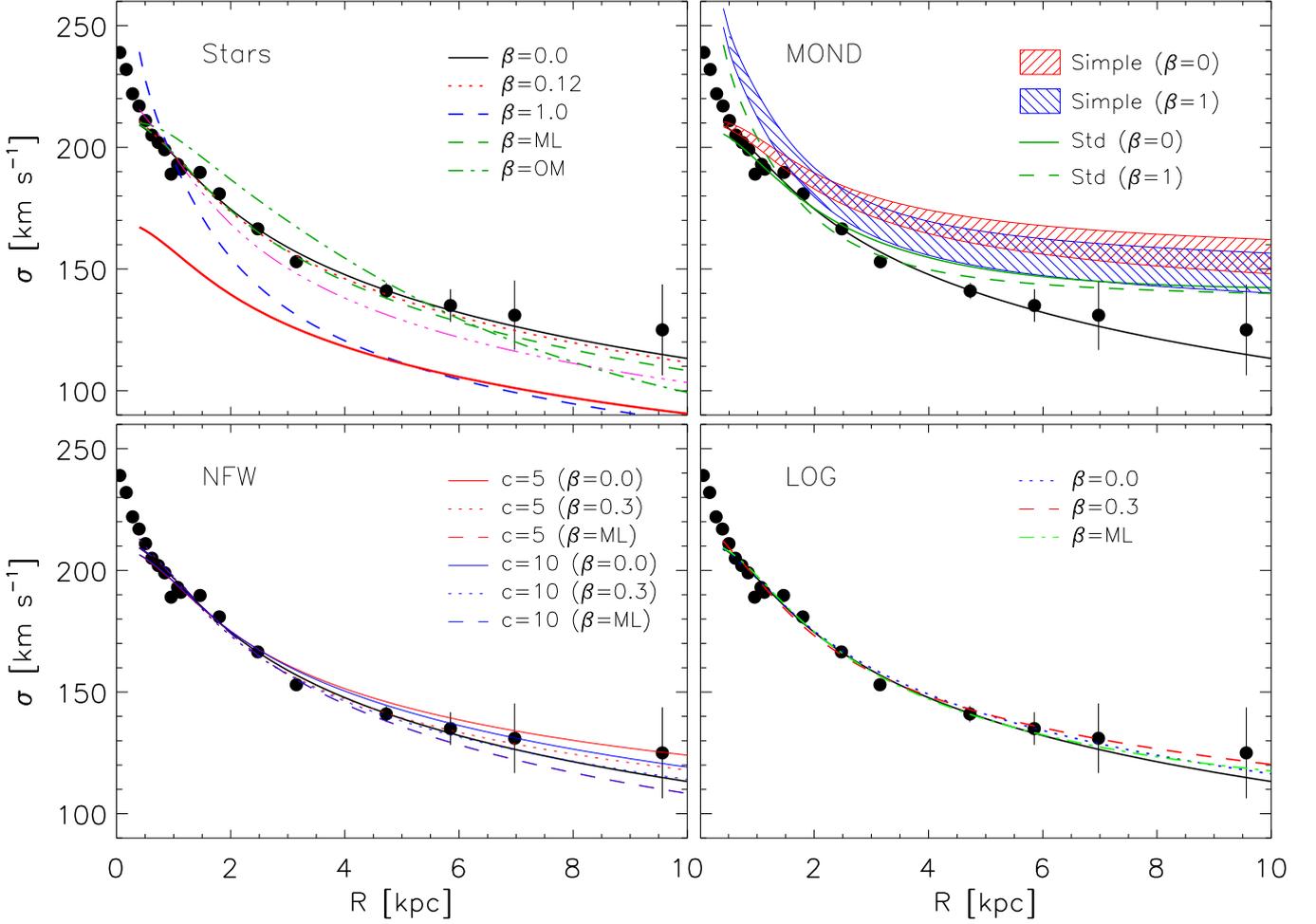}
\caption{Dynamical models of NGC 7507. Black circles indicate in all
  panels the line-of-sight velocity dispersion. All models are given
  outside of 400 pc, where the luminosity model is unaffected by
  seeing. Top left: Stars-only models. The black solid line is an
  isotropic model using $M/L_R$\,=\,3.13. This model is shown in all
  panels as a comparison. The red dotted line is the best-fit
  model for a constant anisotropy using $\beta$\,=\,0.12, while the
  blue dashed line is a fully radial model ($\beta$\,=\,1). The green
  dashed line is the model using the anisotropy as given in
  Eq. \ref{eq:ml}, while the green dot-dashed line is a model using
  the Osipkov-Merrit anisiotropy (Eq. \ref{eq:om}). The red solid line
  is a model with $M/L_R$=2, based on \citeauthor{tortora09}'s stellar
  population analysis (see Section \ref{sec:mlratio}). Top right: MOND
  predictions using the ``simple'' and ``standard'' interpolation
  formulae. The dashed red area is an isotropic model, while the blue
  area indicates a fully radial model for the former
  interpolation. Bottom left: NFW models using $c$\,=\,5 and
  $c$\,=\,10. Parameters for each model are reported in Table
  \ref{tab:nfw_models}. Bottom right: Logarithmic dark halo
  models. Using $\beta$\,=\,0,\, $\beta$\,=\,0.3, and M{\L}
  anisotropy. Parameters for each model are given in Table
  \ref{tab:log_models}.}
              \label{fig:dyn_models}
\end{figure*}

The simplest model is the one in which $M(r)=M_{\star}(r)$ (i.e. no
DM) and $\beta$\,=\,0 (orbital isotropy).  One can adopt the
expected value for the $M/L$ from an stellar populations analysis
(which is a hypothetical value) or treat it as a free parameter in the
Jeans equation solution, which is the procedure we chose here. This
gives an upper limit to the stellar $M/L$.  This model is shown as the
black solid line in Fig. \ref{fig:dyn_models} (top left panel). It
uses an $M/L_R$\,=\,3.13 and already gives a very good description of
the dispersion profile with $\chi^2$\,=\,46.8 for 16 degrees of
freedom, where the statistics is calculated outside 400 pc, which is
the radius where the data are no longer affected by seeing.

When the anisotropy was allowed to vary, we obtained almost the same
$M/L_R$ of 3.11 and $\beta$\,=\,0.12, which is a slightly better fit
(red dotted line). Table \ref{tab:models} shows the parameters for
different best-fit models. An unrealistic totally radial model
($\beta$\,=\,1) is also shown in Fig.~\ref{fig:dyn_models} as a
comparison (blue dashed line).

From theoretical considerations \citep[e.g.][]{abadi06}, elliptical
galaxies are expected to have an anisotropy profile that is
increasingly radially biased with growing radius. An analytical
profile with this characteristic was introduced by M\L05 as
\begin{equation}\label{eq:ml}
\beta_{ML}(r)=\beta_0\frac{r}{r+r_a},
\end{equation}
with $\beta_0$\,$\sim$\,0.5 and $r_a$\,$\sim$\,1.4$R_e$ when fitted to
the merger simulations of \citet{dekel05}. Another profile frequently
used is the Osipkov-Merritt anisotropy model
\citep{osipkov79,merritt85},
\begin{equation}\label{eq:om}
\beta_{OM}(r)=\frac{r^2}{r^2+r_a^2},
\end{equation}
where $r_a$ is a characteristic radius marking the transition to outer
radially biased orbits in both cases.  Since the data do not really
constrain the value for $r_a$, we fixed it to $r_a$\,=\,1.4$R_e$,
using $R_e$\,=\,31\arcsec\, from \citet{faber89}. Even though our
derived effective radius is significantly larger (see
Sect. \ref{sec:light}), we use this older estimation since it has been
used in the previous studies of the galaxy, and also because it
implies a faster transition to a radial behavior. The predictions
from the Jeans equation using these anisotropies are shown in
Fig.~\ref{fig:dyn_models} (top left panel). The M\L05 anisotropy gives
a good description for the dispersion profile (green dashed line), but
formally not better than the constant $\beta=0.12$ model. It reaches a
value of $\beta$\,=\,0.33 at the outermost kinematic point. However,
the Osipkov-Merritt anisotropy (green dash-dotted line) overpredicts
the dispersion for the inner regions and underpredicts it for larger
radii, making it less preferable as a realistic description within the
context of this model.

\begin{table}
  \caption{Comparison of different best-fit dynamical models without DM. Errors indicate the $1-\sigma$ confidence levels.}
  \label{tab:models}
  \centering
  \begin{tabular}{llll}
    \hline\hline
    Model     &  $M/L_{\star,R}$ & $\beta$ & $\chi^2$\\
    &[M$_{\sun}$/L$_{\sun,R}$] & &\\
    \hline
    Stars only & $3.13^{+0.01}_{-0.02}$ & 0.0 &46.8    \\
    Stars only + cst. $\beta$& $3.11^{+0.02}_{-0.03}$ & $0.12_{-0.08}^{+0.08}$& 44.1    \\
    Stars only + M{\L}\tablefootmark{a}&$3.01^{+0.01}_{-0.01}$ &0.33&54.5 \\         
    Stars only + HM\tablefootmark{b}&$2.93^{+0.02}_{-0.02}$&0.45&90.5\\
%\noalign{\smallskip}
    \hline
  \end{tabular}
  \tablefoot{
    \tablefoottext{a}{Indicates the M\L05 anisotropy. The
      value for $\beta$ is the maxiumum value for the anisotropy
      reached at the last kinematic point.}
    \tablefoottext{b}{Indicates the \citet{hansen06} anisotropy modelled using Eq. \ref{eq:ml} with $r_a=500$ pc. The value of $\beta$ has the same meaning as before.}
  }
\end{table}

An interesting relation between anisotropy $\beta$ and the logarithmic
density slope $\alpha$ of a spheroidal system near equilibrium has
emerged from DM simulations. \citet{hansen06} found $\beta(\alpha) = 1
- 1.15~(1+\alpha/6)$, albeit with some scatter. This relation has also
been analyzed by \citet{mamon06} for ellipticals formed in simulations
of mergers of spirals consisting of stars, gas and DM.  They found
different results for the DM, but a quite similar relation for the
stellar density distribution. If this were universal, the dynamical
analysis of spherical systems would be greatly facilitated. We apply
the above relation to NGC 7507 and calculate for our double-beta model
(Eq. \ref{eq:light2}) the logarithmic slope $\alpha =
(r/j)\mathrm{d}j(r)/\mathrm{d}r$.  The resulting anisotropy profile
shows a swift increase from isotropy at the center to $\beta$\,=\,0.3
at around 500 pc, with a mild increase outwards. This behavior can be
reasonably reproduced using the M\L05 anisotropy profile with a small
$r_a$. In the context of a no-DM model, this anisotropy favors an even
lower stellar $M/L$ and underestimates the observed velocity
dispersion at large radii, giving a poor fit to the data.

\subsection{Including a dark halo}
 \label{sec:dark_halos}

Even though the velocity dispersion profile up to $\sim 90\arcsec$ of
NGC 7507 can satisfactorily be described by models without dark
matter, the paradigm of galaxy formation requires the presence of dark
halos. We therefore allow for a (cuspy) NFW profile
\citep[e.g.][]{navarro96} and a (cored) logarithmic halo. In both
cases we fix the stellar $M/L$ to 3.01, therefore maximizing the
contribution of the stellar component to the total mass budget
\citep[``maximum spheroid'', e.g.][]{weijmans09}, this is supported by
the fact that the contribution of DM in the inner regions is thought
to be close to negligible
\citep[e.g.][]{gerhard01,borriello03,mamon05a}.

\subsubsection{NFW models}

\begin{table*}
  \caption{Comparison of different best fitting dynamical models including NFW halos.}
  \label{tab:nfw_models}
  \centering
  %\noalign{\smallskip}
  \begin{tabular}{lcccccccc}
    \hline\hline
    % \noalign{\smallskip}
    $\beta$ & $\rho_s$          & $r_s$  &$r_{200}$ &$M_{200}^{\mathrm{DM}}$    & $M(r_{200})/L_R$ &$f_{\mathrm{DM}} <R_e$ &$f_{\mathrm{DM}} <2R_e$ &$\chi^2$\\
            & [M$_{\sun}$pc$^{-3}$]& [kpc] & [kpc]    & [M$_{\sun}$]&[M$_{\sun}$/L$_{\sun}$] &[\%]& [\%]& \\
    %\noalign{\smallskip}
            \hline
            \multicolumn{7}{c}{Stars ($\Upsilon_R$=3.01) + NFW ($c$=5)}\\ 
            \hline 
            \noalign{\smallskip}
            0.0  &$1.191\times10^{-3}$ & 30 & 150& $3.87^{+3.4}_{-2.1}\times10^{11}$ &7.3 &$3.4^{+0.8}_{-0.9}$ &$7.5^{+2.5}_{-2.5}$ &63.0\\
            0.3  &$1.191\times10^{-3}$ & 25 & 125 & $2.24^{+2.0}_{-1.2}\times10^{11}$ &5.6 &$2.7^{+0.7}_{-0.8}$ &$8.9^{+3.3}_{-3.0}$ &40.5   \\
            M{\L}&$1.191\times10^{-3}$ & 1  & 5  & $1.4^{+1.9}\times10^{7}$     &3.1 &$<1$             & $<1$  &54.5 \\       
            \hline
            \multicolumn{7}{c}{Stars ($\Upsilon_R$=3.01) + NFW ($c$=10)}\\            
            \hline
            0.0 &$6.135\times10^{-3}$  &9& 90 & $8.36^{+6.9}_{-2.5}\times10^{10}$& 4.1 &$3.8^{+1.2}_{-0.5}$&$7.1^{+2.9}_{-1.4}$ &59.3 \\
            0.3 &$6.135\times10^{-3}$  &8& 80 & $5.87^{+2.5}_{-1.9}\times10^{10}$& 3.8 &$3.3^{+0.5}_{-0.6}$&$8.6^{+2.2}_{-1.9}$ &42.9 \\ 
            M{\L}&$6.135\times10^{-3}$ &1& 10 & $0.11^{+7.3}\times10^9$       & 3.1  & $<1$             &$<1$ &54.5 \\
            \hline
         \end{tabular}
         \tablefoot{Column 1 indicates the value for the anisotropy considered for the stellar population. M\L\, indicates the \citet{mamon05b} anisotropy profile with parameters decribed in the text. Column 2 gives the characteristic density of the NFW halo, which comes from the chosen concentration as given in Eq. \ref{eq:nfw_concentration}. Column 3 gives the best fitting scale radius for the chosen concentration.  Columns 4 and 5 give the virial radius and mass of the DM halo, respectively. Masses are also indicated with their 1-$\sigma$ confidence level. Column 6 indicates the virial mass to light ratio. Columns 7 and 8 give the DM fraction within 1 and 2 $R_e$ \citep[using the effective radius given by][]{faber89}, defined as $f_{\mathrm{DM}}$\,=\,$M_{\mathrm{NFW}}/(M_{\mathrm{NFW}} + M_{\star})$, where the baryonic mass is given by Eq. \ref{eq:starmass2}. Column 9 gives the $\chi^2$ for each model.}
     \end{table*}

     The NFW profile is given by
\begin{equation}
\rho_{\mathrm{NFW}}(r)=\frac{\rho_s}{(r/r_s)(1 + r/r_s)^2},
\end{equation}
where $\rho_s$ and $r_s$ are a characteristic density and radius,
respectively; the enclosed mass for this profile is given by
\begin{equation}
M_{\mathrm{NFW}}(r)=4\pi\rho_sr_s^3\left(\ln\left(1+r/r_s\right)-\frac{r/r_s}{1+r/r_s}\right).
\end{equation}

\begin{figure*}[]
\centering
\includegraphics[width=\textwidth]{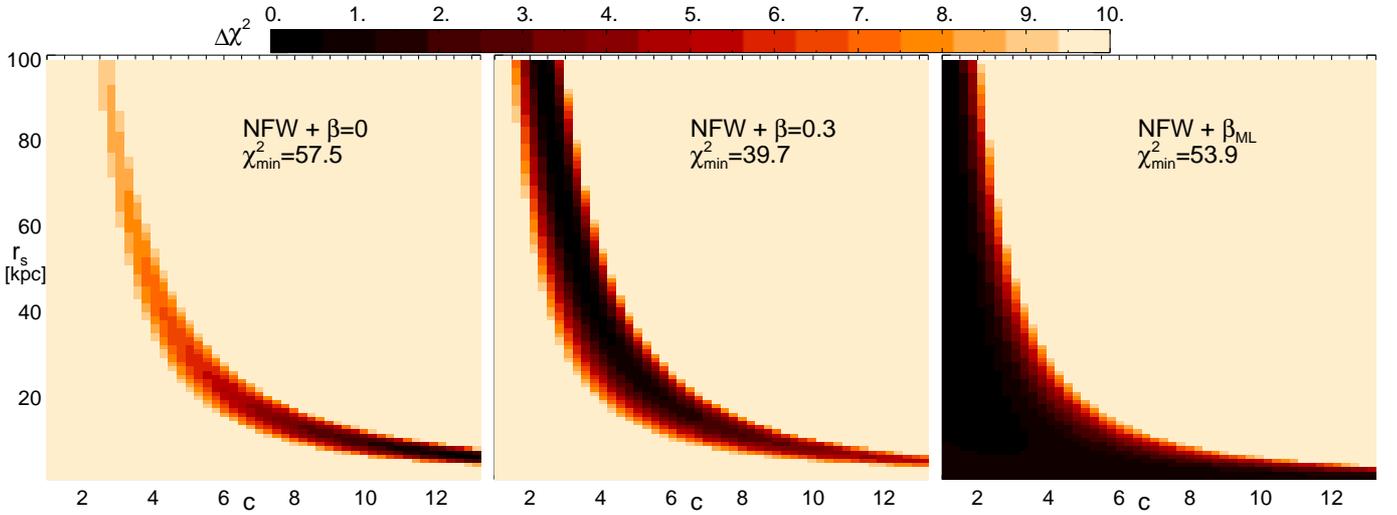}
\caption{$\chi^2$ for models including an NFW halo. Minimum $\chi^2$
  for each model is indicated. Left panel: NFW halo and stars with
  isotropic distribution. Central panel: NFW halo plus stars with a
  radial anisotropy of $\beta$\,=\,0.3. Right panel: NFW halo and
  stars following a M{\L} anisotropy profile.}
              \label{fig:7507:nfw_conf}
\end{figure*}

We explored three models, an isotropic model, a model with a slight
radial anisotropy ($\beta$=0.3) and a model using the M\L\, anisotropy
with $r_a=43\arcsec$ as in Sect. \ref{sec:stars_only}. We excluded the
possibility of tangential orbits beacuse they were inconsistent with
the stars-only models. For each anisotropy, we used a set of
concentrations, $c=\lbrace1,1.25,1.5...13\rbrace$, which encompasses
the range of concentrations found for low luminosity ellipticals
($c\sim5$, \citealt[][]{napolitano08}) and the ones found in
simulations for halos of similar mass \citep[$c\sim10$,
e.g.][]{bullock01,maccio08}. Concentrations are related to the
characteristic density by
\begin{equation}\label{eq:nfw_concentration}
  \frac{\rho_s}{\rho_{\mathrm{crit}}}=\frac{200}{3}\frac{c^3}{\ln(1+c)-c/(1+c)},
\end{equation}
where the critical density is $\rho_{\mathrm{crit}}$\,=\,$3H_0^2/8\pi G$. We
also set a grid of characteristic radii, $r_s=\lbrace
1,2,...,100\rbrace$ kpc, and calculate the $\chi^2$ for each $(c,r_s)$
pair. The results can be seen in Fig. \ref{fig:7507:nfw_conf}. 68\%,
90\% and 99\% confidence levels correspond to
$\Delta\chi^2=\lbrace2.30,4.61,9.21\rbrace$ \citep{avni76}.

As seen in Fig. \ref{fig:7507:nfw_conf}, we cannot constrain
simultaneously $c$ and $r_s$, with the available data/modeling. Even
though we have explored a generous range of parameters, a global
minimum lies beyond the explored space. The minimum $\chi^2$ values
for the isotropic and M{\L} models are worse than the best-fit
stars-only models, even though the models with an NFW profile include
more free parameters. This is an indication that the data shows a
preference for the stars-only models, or more exactly, that our
assumed stellar $M/L_{R}$ is closer to the total $M/L$, which should
be constant across the galaxy. The exception is the model with
$\beta$\,=\,0.3, which gives a significantly smaller $\chi^2$ than all
of the no-DM models.

Since we cannot constrain simultaneously the values for $c$ and $r_s$,
we explore the amount of DM in the NFW halos by selecting the cases
$c$\,=\,5 and $c$\,=\,10. The NFW parameters of each halo and the
assumed concentrations can be seen in Table \ref{tab:nfw_models}. The
velocity dispersion prediction for each of these models is depicted in
Fig. \ref{fig:dyn_models} (bottom left panel).

The best-fit model is the one for $c$\,=\,5 and $\beta$\,=\,0.3, which
gives a virial mass of $2.2\times10^{11}$ M$_{\sun}$. The $c$\,=\,10
model for the same anisotropy gives a slightly worse fit and a
significantly lower virial mass with $5.9\times10^{10}$
M$_{\sun}$. Even though the M\L\, anisotropy should give a more
realistic representation of the true orbital structure of elliptical
galaxies, the mass which correspond to the best-fit scale radius is
very low, although the $1-\sigma$ level is broad with a maximum mass
of $1.9\times10^{10}$ M$_{\sun}$ for $c$\,=\,5. The \citet{hansen06}
anisotropy performs even more poorly, with even lower predicted masses
and much larger $\chi^2$.

The fraction of DM within 1$R_e$ \citep[using the value
from][]{faber89} is lower than 5\% for all models. This number can be
compared with the values found for galaxies of similar brightness
studied by \citet{weijmans09}: NGC 3379 with $f_{DM}(<\!R_e)=0.08$ and
NGC 821 with $f_{DM}(<\!R_e)=0.18$. Even though this low value could
be considered as the natural result of the minimal halo assumption,
this is not exactly the case. If we were to drop this assumption and
allow the stellar mass-to-light ratio to be a free parameter for the
$c=5$ and $c=10$ NFW halos, we would find that the $M/L$ diminishes at
most to a value of 2.9, depending on the assumed anisotropy,
reinforcing the prior assumption. DM fractions within 2$R_e$ can also
be seen in Table \ref{tab:nfw_models}.

\subsubsection{Models with a logarithmic halo}

Even though most, if not all, N-body simulations favor cuspy profiles,
observations in dwarf and low surface brightness galaxies indicate the
presence of constant density cores instead of $r^{-1}$cusps
\citep[see][for a recent review]{deblok10}. From the different types
of cored profiles we selected the logarithmic potential \citep{bt08},
which has a density
\begin{equation}
\rho_{\mathrm{LOG}}(r)=\frac{v_0^2}{4\pi G}\frac{3r_0^2+r^2}{(r_0^2+r^2)^2},
\end{equation}
and enclosed mass
\begin{equation}
M_{\mathrm{LOG}}(r)=\frac{v_0^2r}{G\left(1 + \left(\frac{r_0}{r}\right)^2\right)}.
\end{equation}

The modeling procedure is similar to the one applied in the previous
Section. As for the NFW profiles, the stellar $M/L_R$ remains fixed at
3.01. A grid of parameters was then set with
$r_0=\lbrace1,2...80\rbrace$ kpc and $v_0=\lbrace40,45...215\rbrace$
km s$^{-1}$ and for each pair, the value of $\chi^2$ is calculated
using Eq. \ref{eq:chi2} for the same three anisotropies as in the
previous section; $\beta$\,=\,0, $\beta$\,=\,0.3, and $\beta(r)$
following Eq. \ref{eq:ml}. The grid was later refined for the case
$\beta$\,=\,0. The best-fitting halo parameters for each studied case
are given in Table \ref{tab:log_models}.

The isotropic case allows a light halo with a very small
characteristic radius of only 25 pc. Assuming $\beta$\,=\,0.3 the halo
is much larger with $r_0$\,=\,7 kpc and a circular velocity of 100 km
s$^{-1}$. This radial model provides a significantly tighter fit to
the observed velocity dispersion. This halo is the one that gives the
lowest $\chi^2$ amongst all models (with or without DM). The M\L\,
anisotropy allows a large range of possible halos, but giving a worse
fit than the $\beta$\,=\,0.3 case. As in the case of NFW halos, the
amount of DM enclosed within one $R_e$ is minimal, again having values
of less than 5\% for all models. The velocity dispersion for these
three best-fit models can be seen in Fig. \ref{fig:dyn_models} (bottom
right panel).

  \begin{table}
    \caption{Comparison of different best-fit dynamical models using a logarithmic DM halo. All models use $M/L_{\star,R}=3$.}
         \label{tab:log_models}
        \centering
        \begin{tabular}{cccccc}
            \hline\hline
           % \noalign{\smallskip}
            $\beta$   &  $r_0$ & $v_0$ & $f_{DM} < R_e$ &$f_{DM} < 2R_e$ &$\chi^2$\\
                      &[pc] & km s$^{-1}$& [\%]& [\%]&\\
                      % \noalign{\smallskip}
                      \hline
           % \noalign{\smallskip}
           0.0 & 25 & 60 &4.1 &5.4 &47.7    \\
           0.3 & 7\,000 & 100 &2.3 &7.4 &38.6    \\
           M\L & 74\,000 & 210 & $<1$ &$<1$ &52.4 \\         
            %\noalign{\smallskip}
            \hline
         \end{tabular}
       \end{table}

\subsection{The Gauss-Hermite parameter $h_4$}
\label{sec:h4}

\begin{figure}[t]
\centering
\includegraphics[width=0.50\textwidth]{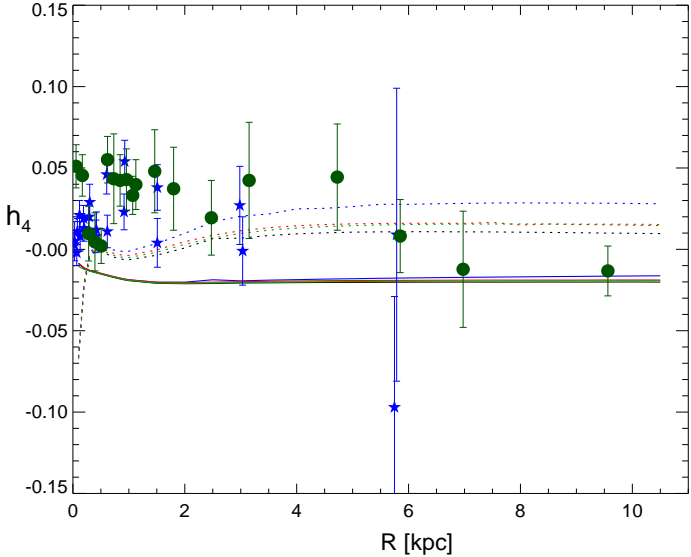}
\caption{Kurtosis models for NGC 7507. Green circles represent our
  $h_4$ measurements as given in Table \ref{tab:dispersions}. Blue
  stars are the measurements from \citetalias{kronawitter00}. Black,
  red, blue, and green lines represent the stars-only, NFW ($c$=5),
  NFW ($c$=10), and logarithmic models, with parameters described in
  the text, respectively. Solid lines are isotropic models, while
  dotted lines are fully radial models.}
              \label{fig:kur_models}
\end{figure}

Predictions for the Gauss-Hermite $h_4$ profile for different models
can be seen in Fig. \ref{fig:kur_models}, where we have followed the
procedure of \citet{lokas02} and related kurtosis and $h_4$ using
\begin{equation}
\kappa(R)\simeq 8\sqrt{6} h_4,
\end{equation}
\citep{vdm93}. We have considered the stars-only, NFW with $c$\,=\,5
and 10, and the logarithmic cases withr isotropy and completely radial
anisotropies. Although the velocity dispersion profiles are best
reproduced with radial anisotropies using $\beta$\,=\,0.3, we did not
calculate their $h_4$ predictions since they would lie between the two
aforementioned models, without adding new information. As in the cases
of the velocity dispersion profiles, since the light profile is
affected by seeing, the model is not expected to reproduce the inner
400 pc.

The measured $h_4$ data are noisy (green circles) and all models
systematically fail to reproduce its high values. We note that the
\citetalias{kronawitter00} data are equally noisy and reach similar
high values (blue stars in Fig. \ref{fig:kur_models}). The difference
is most evident in the inner 500 pc where the
\citetalias{kronawitter00} values are closer to zero and ours scatter
around $\sim$0.05. Although some systematic effect may be shifting our
measurements to higher values, it can hardly be $S/N$ (which is
especially high at small radii) nor template mismatching (innermost
spectra are very well-fitted using the comprehensive MILES stellar
library). Perhaps the only conclusion that we can give is that the
$h_4$ data does not favor isotropic models, which predict slightly
negative values for $h_4$ (see Sect. \ref{sec:limitations} for an
alternative explanation).

%%%%%%%%%%%%%%%%%%%%%%%%%%%%%%%%%%%%%%%%%%%%%%%%%%%%%%%%%%%%%%%%%%%%%%%%
\section{Discussion}
\label{sec:disc}
\subsection{Stellar $M/L$}
\label{sec:mlratio}

The stellar $M/L$ determines the amount of baryonic matter and thus,
based on its difference from the total mass, also the fraction of dark
matter.  Our $M/L$-value is not quite consistent with what one would
theoretically expect for an old, metal-rich population by comparison
with published population synthesis models, but it is in agreement
with literature values.

\citetalias{kronawitter00} gives a central $M/L_B$ value of 5.9 for a
distance of 26.9 Mpc. Since dynamically derived $M/L$s are inversely
proportional to distance, at our adopted distance of 23.2 Mpc this
value transforms into $M/L_B$\,=\,6.8, which gives $M/L_R$\,=\,3.14
using $(B\!-\!R)$\,=\,1.90 \citep{franx89a} and
$(B\!-\!R)_{\sun}$\,=\,1.06 \citep{bm98}, in excellent agreement with
our value. \citet{magorrian01}, based on kinematic data by
\citet{bertin94}, found $M/L_B\!\sim\!7$, somewhat higher than our
value, but also in agreement considering their shorter distance (21
Mpc).

The stellar $M/L$s even of old elliptical galaxies are apparently not
universal: \citet{cappellari06} find values in the range 1.5 $\lesssim
M/L_I \lesssim$ 6.0. For NGC 7507, $M/L_R$\,=\,3.1 translates into
$M/L_I$\,=\,2.42. If we only consider elliptical galaxies in the
\citet{cappellari06} sample, that is, excluding S0 galaxies, this
range would narrow to 2.3$\lesssim M/L_I \lesssim$6.1, putting NGC
7507 among the ellipticals with the lowest $M/L$.  This value, unless
the metal abundance of NGC 7507 is distinctly subsolar, would indicate
a somewhat younger galaxy, with an age around 8 Gyr for a solar
abundance, which agrees with previous line-strength measurements of
this galaxy \citep{trager98,ogando08}. This is unsurprising because a
large fraction of isolated elliptical galaxies are known to be of
younger age \citep{collobert06,reda07}, which is consistent also with
simulations \citep{niemi10}. \citeauthor{tortora09}'s (2009) stellar
population models give a very similar age of 7.7 Gyr (Kroupa IMF), but
a siginificantly different $M/L_R=2$. This value is in contradiction
with the line strength analyses and with our dynamically derived $M/L$
and would imply a huge DM content, even in the innermost
parts of the galaxy (red solid line in Fig. \ref{fig:dyn_models}, top
left panel).

\subsection{Dark matter?}

That an isotropic model with no DM closely fits the observations
nicely, does not imply that there is no DM. This situation
was realized before in analyses where PNe were used as dynamical
tracers \citep[][M\L05]{romanowsky03,dekel05}. However, the relatively
low stellar $M/L$-value of NGC 7507 is in line with the finding that,
regardless of the situation at larger radii, the DM content
in the inner region of elliptical galaxies is negligible
\citep[][]{gerhard01,mamon05a,cappellari06}. Within the framework of a
spherical model, a possibility to host DM at larger radii is
a radial anisotropy. It must then be a coincidence that the observed
dispersion so closely follows the predictions of the photometric model
under isotropy. This is a difference to the use of PNe as dynamical
tracers, where the density profile of the parent population is
notoriously uncertain.

Are the stellar orbits of NGC 7507 affected by a strong radial
anisotropy? The comparison with the kurtosis profiles for a few models
shows that a fully radial anisotropy of the stellar component can
produce a consistently positive kurtosis of the observed order,
although failing in the innermost regions.  However, we say that with
caution, since the kurtosis might also be influenced by rotation
\citep{dekel05}.

NGC 7507 with a stellar mass of about $2\times10^{11}$ M$_{\sun}$
belongs to the most massive isolated elliptical galaxies in the
simulations of \citet{niemi10}. The dark halos in which they are
embedded have predicted masses of the order of a few times $10^{12}$
M$_{\sun}$ within their virial radii. Our halos from Table
\ref{tab:nfw_models} are far from this mass, although we caution that
the extrapolation to the virial radius might be uncertain
\citep{mamon05b}.

\subsection{Limitations of a spherical model}
\label{sec:limitations}
Even though the apparent spherical symmetry of the galaxy and the
small amount of rotation give reasons to use our modeling approach,
at the same time other observations may point to a more complex
scenario.  High $h_4$ values could be signature of an undetected
face-on disk or flattening along the line of sight
\citep{magorrian99,magorrian01}, which might also explain the failure
to reproduce the $h_4$ values when using the \citet{lokas02}
simplification. The second observed particularity is the rotation
along the minor axis. Elliptical galaxies with this property are rare:
\citet{franx89a} found that minor axis rotation was larger than major
axis rotation in only 2 out of 22 galaxies in their sample. This
behavior is a signature of triaxial galaxies
\citep[e.g.][]{binney85}. Another sign of triaxiality comes from the
outermost velocity along the major axis which shows a significant
difference with the inner measurements. This change in the mean
velocity corresponds to a sudden change in the ellipticity at the same
radius ($\sim$100\arcsec, Figs. \ref{fig:photometry} and
\ref{fig:major}).

\subsection{MOND and the baryonic Tully-Fisher relation}
\label{sec:mond}

Among the problems that MOND faces is how the Newtonian and MONDian
regimes are linked. Taking this interpolation function as
$\mu(x)=x/(1+x)$ \citep[known as ``simple'',][]{famaey05}, the MONDian
circular velocity is
\begin{equation}
  \varv_{\mathrm{circ,MOND}}^2(r)=\frac{\varv^2_{\mathrm{circ,N}}}{2}+\sqrt{\frac{\varv^4_{\mathrm{circ,N}}}{4} +a_0\varv^2_{\mathrm{circ,N}}r}
\end{equation}
\citep[e.g.][]{richtler08}, where $\varv_{\mathrm{circ,N}}$ is the Newtonian
circular velocity associated with the stellar component and
$a_0=1.35^{+0.28}_{-0.42}\times10^{-8}$cm\,s$^{-2}$ \citep{famaey07} is the
acceleration constant that separates the MOND and Newtonian realms. An
alternative interpolation function, dubbed ``standard'', is
$\mu(x)=x/\sqrt{1+x^2}$ \citep{sanders02}, for which the circular
velocity is given by
\begin{equation}
  \varv_{\mathrm{circ,MOND}}^4(r)=\frac{\varv^4_{\mathrm{circ,N}}}{2}+\sqrt{\frac{\varv^8_{\mathrm{circ,N}}}{4} +a_0^2\varv^4_{\mathrm{circ,N}}r^2}
\end{equation}
\citep[e.g.][]{samurovic08}. The masses associated with these
circular velocities are introduced in Eq. \ref{eq:jeans4} to obtain
the MOND predictions for the projected velocity dispersion profile for
the isotropic case, using in all cases $M/L_{\star,R}=3$. As seen in
Fig. \ref{fig:dyn_models} (top right panel), the simple interpolation
formula cannot explain the velocity dispersion profile under
isotropy. The outermost measured points can only be explained under a
totally radial model and only when the lowest allowed values for $a_0$
are considered. The standard interpolation formula (green lines in the
same plot) gives very similar predictions for the isotropic and radial
cases for the outermost bins, and again favors a lower value for
$a_0$.

The barred spiral galaxy NGC 7513 is found at an angular distance of
18.2 arcmin from NGC 7507. In the MOND context, it is interesting to
ask whether it could provide an external gravitational field which
would modify the MOND phenomenology in NGC 7507. The mean of the
published Tully-Fisher distances is 19.3 Mpc and thus places NGC 7513
with respect to NGC 7507 (23 Mpc) somewhat in the foreground and a
separation of 4 Mpc would produce a very weak field.  Adopting an
inclination-corrected HI-line width of 275 km s$^{-1}$ as the value
for a constant MONDian rotation curve \citep{springob07}, the
acceleration at NGC 7507 would be $6.3\times10^{-13}$ m s$^{-2}$.
Assuming the same distance as for NGC 7507, the separation would be
123 kpc and the acceleration at NGC 7507 would be $2\times10^{-11}$ m
s$^{-2}$. Therefore, an external field should not influence any of the
MONDian dynamics of NGC 7507.

The baryonic Tully-Fisher relation (BTFR) of spiral galaxies relates
circular velocity (which is a proxy for the \textit{total} mass) to
the \textit{baryonic} content of the galaxy. This relation, which
reads $M_{\mathrm{bar}}$ = 50 $\varv_{f}^4$, where $\varv_{f}$ is the
circular velocity in the flat part of the rotation curve, has been
found to be correct over many orders of magnitude
\citep{mcgaugh05,trachternach09}. A tight relation between baryonic
mass and circular velocities is not expected when DM is
assumed to dominate the dynamics, but it finds a natural explanation
under MOND \citep{sanders02}. Similarly to spirals, early-type
galaxiess have been found to follow a relation that is analogous to
the Tully-Fisher relation \citep{gerhard01,magorrian01}, but since
circular velocities up to a large radius are difficult to obtain,
there is no surprise in finding that ellipticals follow a relation
that is offset from the spiral BTFR, i.e. circular velocities are
probably overestimated. For NGC 7507, we calculated the circular
velocity at the outermost point with measured kinematics of 250 km
s$^{-1}$. Using \citeauthor{mcgaugh05}'s version of the BTFR, this
implies a mass of 2$\times10^{11}$ M$_{\sun}$. Using
Eq. \ref{eq:starmass2} with our preferred $M/L$ value, we obtained
2.5$\times10^{11}$ M$_{\sun}$, that is, NGC 7507 would lie somewhat
above the BTFR for spirals; this appears inconsistent with previous
results for other ellipticals \citep{gerhard01}, although consistent
when all uncertainties are taken into account. We also note that the
slope of 4 required by MOND remains a disputed value \citep[see
e.g.][]{gurovich10}.

\section{Summary and conclusions}
\label{sec:conclusions}
We have obtained wide-field photometry in Kron-Cousins $R$ and
Washington $C$ as well as new long-slit spectroscopy of the field
elliptical NGC 7507 out to about $\sim 90\arcsec$, reaching farther
out than previous studies.

We have measured almost no rotation along the major axis and
significant rotation ($\Delta \varv\sim 50$ km\,s$^{-1}$) along the
minor axis (although the galaxy is almost perfectly circular). The
velocity dispersion profile of the galaxy shows a rapid decline along
both axes.

We have performed a spherical Jeans analysis to find the mass profile
that most closely represents the projected velocity dispersion
profile. When assuming isotropy, a radially constant $M/L_R$ ratio of
3.1 would be sufficient to reproduce the velocity dispersions. This
value corresponds, if solar abundance is assumed, to an age in the
range 8-10 Gyrs.  DM halos can be included in the isotropic case, but
provide significantly poorer fits than the stars-only models. When we
allowed for radial anisotropies and cored/cuspy dark halos, we were
able to improve the fit to the velocity dispersion profile
marginally. This contradicts the result od \citet{kronawitter00} and
\citet{gerhard01} who found this galaxy to be one of the best examples
for hosting a dark halo. It is however in line with the results of
\citet{magorrian01} who also found no clear evidence of DM.

The most massive NFW halo, using $\beta$\,=\,0.3 and a concentration
parameter of 5, has a scale $r_s$\,=\,25 kpc, implying a virial mass
of only $2.2^{+2.1}_{-1.2}\times10^{11}$ M$_{\sun}$. Modified
Newtonian dynamics (MOND), if applied straightforwardly, predicts
velocity dispersions that are too high.

It appears that NGC 7507 is a very interesting case in the discussion
of DM in elliptical galaxies. It remains to be seen whether
non-spherical models based on an extended data set, perhaps including
also PNe and/or globular clusters, would provide different results.

%%%%%%%%%%%%%%%%%%%%%%%%%%%%%%%%%%%%%%%%%%%%%%%%%%%%%%%%%%%%%%%%%%%%%%%%
\begin{acknowledgements}
  We thank the referee, Dr. Gary Mamon, for a careful and insightful
  report. We thank Roberto Saglia for providing the kinematic
  measurements from \citet{kronawitter00} in electronic form. We thank
  Srdjan Samurovi\'c and Michael Hilker for making available their
  Linux compatible versions of R. van der Marel's code; Michele
  Cappellari and Anne-Marie Weijmans for useful advice on the usage of
  pPXF. RS acknowledges support from a CONICYT doctoral fellowship and
  the ESO Studentship Program. TR acknowledges support from the
  Chilean Center for Astrophysics, FONDAP Nr. 15010003, and from
  FONDECYT project Nr. 1100620. LPB acknowledges support from CONICET,
  Agencia Nacional de Promoci\'on Cient\'ifica y Tecnol\'ogica, and
  Universidad Nacional de La Plata (Argentina). AJR was supported by
  National Science Foundation grants AST-0808099 and AST-0909237.
\end{acknowledgements}

\bibliographystyle{aa}
\bibliography{7507}
\Online

\end{document}